\documentclass[conference,10pt]{IEEEtran}
\usepackage[utf8]{inputenc}
\IEEEoverridecommandlockouts
\usepackage{amsmath,amssymb,amsthm}
\usepackage{enumerate}
\usepackage{stfloats}
\usepackage{comment}
\usepackage{bbm}
\usepackage{siunitx}
\usepackage{mathtools}
\usepackage{accents,latexsym,cancel}
\usepackage{cite}

\usepackage[dvipsnames]{xcolor}
\definecolor{myPink}{RGB}{255,105,183}

\usepackage[T1]{fontenc}
\usepackage{graphics} 
\usepackage{epsfig} 
\usepackage[mathscr]{euscript}
\usepackage{algorithm}
\usepackage[noend]{algpseudocode}
\usepackage{bbm}
\makeatletter
\def\BState{\State\hskip-\ALG@thistlm}
\makeatother

\usepackage{tikz}
\usetikzlibrary{arrows,shapes,chains,matrix,positioning,scopes,patterns,calc,
decorations.markings,
decorations.pathmorphing,
}

\usepackage{pgfplots}
\pgfplotsset{compat=1.3}
\usepgflibrary{shapes}

\renewcommand{\epsilon}{\varepsilon}

\newcommand{\RNum}[1]{\uppercase\expandafter{\romannumeral #1\relax}}

\newcommand{\av}{\ensuremath{\mathbf{a}}}

\newcommand{\cv}{\ensuremath{\mathbf{c}}}

\newcommand{\gv}{\ensuremath{\mathbf{g}}}

\newcommand{\mv}{\ensuremath{\mathbf{m}}}
\newcommand{\nv}{\ensuremath{\mathbf{n}}}

\newcommand{\xv}{\ensuremath{\mathbf{x}}}
\newcommand{\yv}{\ensuremath{\mathbf{y}}}
\newcommand{\zv}{\ensuremath{\mathbf{z}}}

\newcommand{\Am}{\ensuremath{\mathbf{A}}}

\DeclareMathAlphabet{\mcl}{OMS}{cmsy}{m}{n}

\newlength\tikzwidth
\newlength\tikzheight

\definecolor{mycolor1}{rgb}{0.63529,0.07843,0.18431}%
\definecolor{mycolor2}{rgb}{0.00000,0.44706,0.74118}%
\definecolor{mycolor3}{rgb}{0.00000,0.49804,0.00000}%
\definecolor{mycolor4}{rgb}{0.87059,0.49020,0.00000}%
\definecolor{mycolor5}{rgb}{0.00000,0.44700,0.74100}%
\definecolor{mycolor6}{rgb}{0.74902,0.00000,0.74902}%

{\begin{list}{$\bullet$}
   {\setlength{\itemsep}{0in}
    \setlength{\labelsep}{0.05in}
    \setlength{\labelwidth}{0.2in}
    \setlength{\leftmargin}{0.25in}
    \setlength{\rightmargin}{0in}
    \setlength{\topsep}{0in}
   }}
{\end{list}}
\title{PolarAir: A Compressed Sensing  Scheme for Over-the-Air Federated Learning}
\author{Michail Gkagkos, Krishna R. Narayanan,  Jean-Francois Chamberland, Costas N. Georghiades \\
Department of Electrical and Computer Engineering, Texas A\&M University
\thanks{
This material is based upon work supported, in part, by the National Science Foundation (NSF) under Grants CCF-2131106 \& CNS-2148354, and by Qualcomm Technologies, Inc., through their University Relations Program.}
}

\usepackage[numbers]{natbib}
\usepackage{natbib}
\usepackage{graphicx}
\usepackage{algorithm}
\newcommand{\T}{^{\mbox{\tiny T}}}
\newcommand{\pd}{\text{P}_{\mathrm{d}}}
\newcommand{\pfa}{\text{P}_{\mathrm{fa}}}
\begin{document}

\maketitle

\begin{abstract}
We explore a scheme that enables the training of a deep neural network in a Federated Learning configuration over an additive white Gaussian noise channel.
The goal is to create a low complexity, linear compression strategy, called PolarAir, that reduces the size of the gradient at the user side to lower the number of channel uses needed to transmit it.
The suggested approach belongs to the family of compressed sensing techniques, yet it constructs the sensing matrix and the recovery procedure using multiple access techniques.
Simulations show that it can reduce the number of channel uses by $\sim$30\% when compared to conveying the gradient without compression.
The main advantage of the proposed scheme over other schemes in the literature is its low time complexity. 
We also investigate the behavior of gradient updates and the performance of PolarAir throughout the training process to obtain insight on how best to construct this compression scheme based on compressed sensing.
\end{abstract}

\section{Introduction}



\subsection{Federated Learning and Over-the-Air Federated Learning}
Federated Learning (FL) was first introduced as Federated Optimization in \cite{FOptimization-google}.
The main idea is that the workers use their available data to solve a Machine Learning (ML) task and the Parameter Server (PS) orchestrates the system by sending the current model to workers.
In addition, there are some characteristics that describe FL.
For example, the samples on each worker can be drawn from a different distribution and the number of available data at each worker may differ by orders of magnitude, known as non-i.i.d.\ and unbalanced datasets, respectively.
The term Federated Learning was established in \cite{googleFL2016,googleFL2017}, and FedAvg \cite{FedAvg} is the original FL algorithm.
The basic steps of FL are as follows.
The PS selects a fraction of the workers in the system to join the training and broadcasts the current model.
Then, the workers use their available data to calculate local model updates.
The average of the updated models is computed by the PS, after receiving messages from the participating workers.
The information sharing step is called the communication round.
The ML training task goes back and forth between the workers and the PS; training terminates after a certain accuracy is achieved.
The main bottleneck in this framework is the communication cost.
To alleviate it, the algorithm in~\cite{FedAvg} has workers update their models locally for $E$ epochs before sending their updated models to the PS.
Alternatively, in the FedPAQ scheme introduced in \cite{Hassani-FedPAQ}, local updates and quantization are applied to lower the cost.
A majority of the algorithms in FL assume that users communicate to the PS using a medium access control (MAC) layer protocol that prevents interference between users. 
Therefore, each user in the network can use digital coding and modulation techniques to reliably communicate the compressed quantized gradients over a noisy channel.
However, existing literature related to distributed sensing has shown that the superposition nature of the wireless medium can be exploited to directly compute functions (in particular, the sum over the complex field) of each user's modulated data \cite{sensorOver,GastparCompuMAC,analogFunctions,wilson2010joint}.

The superposition property of the wireless channel has been exploited by Amiri and Gunduz for computing the sum of gradients in FL under the name of over-the-air computation in \cite{Amiri-AMP-AWGN}.
This scheme sparsifies the gradient and then applies compressed sensing (CS) techniques to create messages.
A scaled version of the sketch is transmitted over a synchronous multiple access AWGN channel, and a noisy sum of the sketches is received at the PS.
 We refer to the received signal as the measurement vector. 
In this scheme, Approximate Message Passing (AMP) is used to recover the sum of the gradients from the measurement vector.
They call this algorithm an analog computation scheme since the transmitted values are unquantized real  numbers.
Amiri and Gunduz show that the analog scheme in \cite{Amiri-AMP-AWGN} can achieve better performance than the digital schemes in SignSGD \cite{SIGNSGD}, QSGD \cite{QSGD}.
Similar ideas have been applied to Single Input Single Output (SISO) fading channels \cite{AmiriFadingChannels}, and to Multiple Input Multiple Output (MIMO) channels \cite{Amiri-CS-MassiveMIMO}.
More recent algorithms in this area include those described in \cite{RLC,Accelerated,Temporal,amiriMIMO-MAC,Over-the-AirStatistical} and the references therein.

\subsection{Main Contributions}

In the three schemes found in \cite{Amiri-AMP-AWGN,AmiriFadingChannels,Amiri-CS-MassiveMIMO}, the authors aim to train a Neural Network (NN) to classify images of the MNIST dataset.
Since the classification of this dataset can be achieved by a NN with a small number of parameters, AMP can be used to recover the sum of sparse vectors.
In the present article, we consider the task of classifying images of the CIFAR-10 dataset \cite{cifar}.
This ML task is more challenging and, to reach this goal, we use one of the Residual Networks \cite{ResNets,ResNet20}, which contains around 270,000 parameters.
Since the time complexity of AMP scales linearly with the length of the sparse vector, a low-complexity alternative to AMP is desirable.
Hence, we propose an algorithm, called PolarAir, that combines $K$-sparsification and CS, in a manner akin to \cite{asitPolarScheme} and \cite{gkagkosISIT2021}, to train the ResNet20 \cite{ResNet20} over the AWGN medium.
The time complexity of the proposed scheme, and the scheme in \cite{Amiri-AMP-AWGN}, is $\mathcal{O}(K^3 + K^2\log N)$ and $\mathcal{O}(MN)$, respectively,  where  $N$ is the length of the gradient, and $K \ll N$ is sparsity and $M$ is the number of measurements.
Also, we examine the performance of the CS scheme during the training phase.
Our findings provide new insights on how to construct CS algorithms for Over-the-Air FL.

\section{System Model}
We consider a wireless network with $W$ mobile devices called workers, and one remote PS.
Workers communicate only with the PS.
We denote the collection of all available data in the network by $\mathbf{D}$, and the set of available data to worker~$w$ by $\mathbf{D}_w \subset \mathbf{D}$.
The goal is to train a ML model using distributed data available at each worker node~$w$. 
This is accomplished by having workers communicate gradients to the PS sequentially, through several update rounds. 
Each communication round consists of three operations.
First the PS broadcasts the parameters of the current  model $\boldsymbol\theta_t$ to all the workers through a reliable communication scheme.
Note that this assumption has appeared in previously publish work, e.g., \cite{Amiri-AMP-AWGN}.
Each worker then employs a subset of their locally available data $B_w(t) \subset \mathbf{D}_w$ (also known as a batch) together with the current model to compute a gradient $\gv_w(\boldsymbol\theta_t)$.
This gradient is then compressed and encoded into the signal $\xv_w(t)$, which is transmitted to the PS using the (synchronous) multiple access AWGN channel.
The signal received at the PS during round~$t$ is
\vspace{-5mm}
\begin{equation} \label{eq:RXsignal}
\yv(t) = \sum_{w=1}^W\xv_w(t) + \zv(t)
\end{equation}
where $\xv_w(t) \in \mathbb{R}^{M}$ is the channel input transmitted by worker $w$, which is a function of the gradient, $\yv(t) \in \mathbb{R}^{M}$ is the received signal and $\zv(t)$ is the AWGN vector with independent and identically distributed (i.i.d.) $N(0,1)$ entries.
After the training step, the updated model is multicasted to the workers by the PS through a noiseless medium.

The recovery of the average of the $W$ gradients, i.e., $\frac{1}{W}\sum_{w=1}^W\gv_w(\boldsymbol\theta_t)$, can be viewed as a computation problem over a MAC \cite{GastparCompuMAC}; this problem has been studied in the sensor networks literature \cite{analogFunctions}, \cite{sensorOver}.
In the present context, our goal is to minimize the length of the vector $\xv_w(t)$ that can be used to transmit the gradient, without losing in terms of the test accuracy.
To achieve that, we apply a top-$K$ sparsification in conjunction with a CS-based signaling scheme similar to \cite{asitPolarScheme,gkagkosISIT2021}.
We focus on an elementary channel model to highlight the main idea behind PolarAir.
Nevertheless, it is possible to create extensions of the proposed scheme suited to more intricate channel models (see, e.g., \cite{AmiriFadingChannels}).

\section{PolarAir}
The proposed scheme borrows ides from \cite{Amiri-AMP-AWGN}, and the main contribution of this paper is the CS scheme described in Section \ref{sc:CSalgo}.
The main ideas are a top-$K$ sparsification step, which increases the amount of zeros in the gradient, and a CS encoding step that captures the sparse gradient by taking linear measurements.
The first component of the compression algorithm is motivated by the fact that sparse updates act in a manner similar to a regularization term.
Then, leveraging CS techniques make sense because a $K$-sparse vector of length $N$ can be stored by taking
 $O(K\log\frac{N}{K})$ linear measurements \cite{eldar2012compressed}.
Since the map from the sparse vector to its measurement vector is linear, the sum of the sparse gradients can be recovered from the sum of their corresponding measurement vectors, under suitable conditions.
More specifically, our goal is to recover the largest $K$ values of the sum of the $W$ transmitted $K$-sparse vectors.
In this paper, ``$K$ largest values'' means the $K$ largest values in magnitude.
Also, we note that each worker has a different dataset and, hence, we expect the sum of $W$ gradient vectors, each with at most $K$ non-zero entries, to yield an aggregate vector with at most $KW$ non-zero entries.
Still, we only wish to recover the strongest $K$ values in this group.
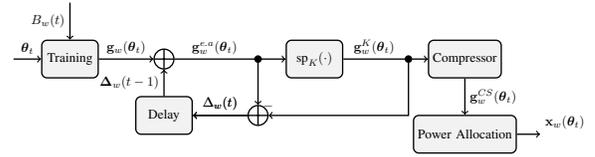
\begin{figure}
    \centering
 \scalebox{0.5}{\definecolor{myColor}{RGB}{36, 173, 71}
\begin{tikzpicture}
  [
  font=\normalsize, draw=black, line width=1pt,
  block/.style={rectangle, minimum height=10mm, minimum width=15mm,
  draw=black, fill=gray!10, rounded corners},
  sum/.style={rectangle, minimum height=15mm, minimum width=15mm, draw=black, rounded corners, fill=gray!10},
  message/.style={rectangle, minimum height=5.5mm, minimum width=80mm, draw=black, rounded corners},
  submessage/.style={rectangle, minimum height=5.5mm, minimum width=40mm, draw=black, rounded corners},
  inputCh/.style={rectangle, minimum height=5.5mm, minimum width=20mm, draw=black, rounded corners}
  ]
  \tikzstyle{sum}   = [circle, minimum width=15pt, draw, inner sep=0pt, path picture={\draw (path picture bounding box.south) -- (path picture bounding box.north) (path picture bounding box.west) -- (path picture bounding box.east);}]

\node[block,align=center] (train)  {Training};
\draw[->] (-1.5,0) -- (train.west) node[midway,above]{$\boldsymbol\theta_t$};
\draw[->] (0,1.5) -- (train.north) node[midway,left]{$B_w(t)$};

 \node [sum, right of=train, node distance = 2.5cm] (sum) {};
\node[circle,draw, right of=sum,text=white,fill=black, scale=0.5,node distance = 2.5cm] (c1) {};
 \node[block,below of= sum, node distance = 1.5cm] (delay) {Delay};
  \draw[->] (delay) -- (sum)node[midway,left]{$\boldsymbol\Delta_w(t-1)$};
 \draw[-] (sum) -- (c1)node[midway,above]{$\gv_w^{e.a}(\boldsymbol\theta_t)$};
  \draw[->] (train) -- (sum)node[midway,above]{$\gv_w(\boldsymbol\theta_t)$};
  
\node[block,right of= sum, node distance = 4cm] (sparse) {$\text{sp}_K(\cdot)$};
 \node[circle,draw, right of=sparse,text=white,fill=black, scale=0.5,node distance = 2.5cm] (c2) {};
  \draw[-] (sparse) -- (c2)node[midway,above]{$\gv_w^{K}(\boldsymbol\theta_t)$};
  \draw[->] (c1) -- (sparse);
  
  \node [sum, below of= c1, node distance = 1.5cm] (subtract) {};
 \draw[->] (subtract) -- (delay)node[midway,above]{$\Delta_w(t)$};
  \draw[->] (c1) -- (subtract);
  \draw[->] (subtract) -- (delay)node[midway,above]{$\Delta_w(t)$};
  \draw[->] (c2) |- (subtract)node[left,above right]{$-$};

\node[block,right of= c2, node distance = 1.5cm] (CS) {Compressor};
\draw[->] (c2) -- (CS);

\node[block,below of= CS, node distance = 2cm] (power) {Power Allocation};
\draw[->] (power.east) -- (12.5,-2) node[left,above right]{$\xv_w(\boldsymbol\theta_t)$};

\draw[->] (CS.south) -- (power.north) node[midway,right]{$\gv_w^{CS}(\boldsymbol\theta_t)$};

\end{tikzpicture}}
        \caption{The protocol all workers follow; similar to the architecture in \cite{Amiri-AMP-AWGN}}
        \label{fg:workersBD}
\end{figure}

\subsection{Worker}

All workers involved in the training process follow the same protocol for every communication round.
We restrict our attention to worker~$w$ and communication round~$t$ to describe the steps of PolarAir.
A block diagram of the procedure at the worker side is given in Fig.~\ref{fg:workersBD}.
First, the worker receives the model parameters $\boldsymbol\theta_t$ from the PS.
Then, based on the batch size, $|B_w(t)| \leq |\mathbf{D}_w|$, worker~$w$ trains the current model; that is, it computes the gradient $\gv_w(\boldsymbol\theta_t)$.
To keep the estimator unbiased, the local accumulation error is added to the gradient and the resulting vector becomes
\begin{align}
    \gv_w^{e.a.}(\boldsymbol\theta_t) = \gv_w(\boldsymbol\theta_t) + \boldsymbol\Delta_w(t-1),
\end{align}
where $\boldsymbol\Delta_w(t-1)$ is the error defined in \eqref{eq:error}, with initial condition $\boldsymbol\Delta_w(0) = \mathbf{0}$ at time $t=0$.
Furthermore, a top-$K$ sparsification operation is applied, and the resulting vector of length $N$ has only $K$ non-zero entries.
We denote the top-$K$ sparsification function as $\operatorname{sp}_K(\cdot)$ and the input and output of this function are indicated by $\gv_w^{e.a.}(\boldsymbol{\theta}_t)$ and $\gv_w^{\text{\tiny{K}}}(\boldsymbol\theta)$, respectively.
Note that $K$ is fixed during the training process, and it is known to all participants a priori.
The next step is to compute the error between the dense vector $ \gv_w^{e.a.}(\boldsymbol\theta_t)$ and the sparsifing vector $\gv_w^{\text{\tiny{K}}}(\boldsymbol\theta_t)$.
The error is indicated by $\boldsymbol{\Delta}_w(t)$ and can be computed as 
\begin{align}
    \boldsymbol\Delta_w(t) = \gv_w^{e.a.}(\boldsymbol\theta_t) - \gv_w^{\text{\tiny{K}}}(\boldsymbol\theta_t) \label{eq:error}
\end{align}
Next, a CS scheme, similar to \cite{gkagkosISIT2021} and \cite{asitPolarScheme}, is used to reduce the dimension of vector $\gv_w^{\text{\tiny{K}}}(\boldsymbol\theta_t)$ from $N$ to $M_t < N$.
For the time being, we treat the CS algorithm as a black box; the details of the adopted scheme are available in Section~\ref{sc:CSalgo}.
We denote the output of the CS encoding by $\gv_w^\text{\tiny{CS}}(\boldsymbol\theta_t) \in \mathbb{R}^{m_t}$.
Because the values of $\gv_w^\text{\tiny{CS}}(\boldsymbol\theta_t)$ are unconstrained, the resulting vector does not satisfy a power constraint.
Thus, we apply power control in a manner akin to the Mean-Removal power control of \cite{Amiri-AMP-AWGN}.
A main difference between our CS scheme and the scheme in \cite{Amiri-AMP-AWGN} is that the former is adaptive.
In other words, the number of measurements increases as the performance of the recovery algorithm decreases.
We explain when and how to increase $m_t$ in Section~\ref{sc:expe}.
The average power of the symbol is set to $P$ and, therefore, the power of the channel input $\xv_w(t)$ is linear in the number of measurements.
Particularly, the power of the transmitted signal is $
   \Vert \xv_w(t) \Vert_2^2= P(m_t+2)=  P{M}_t$.
By establishing the power constraint of the problem, we can proceed with the power control algorithm.
The first step of algorithm in \cite{Amiri-AMP-AWGN} is to compute the mean of the measurement vector, i.e., $\mu_w(t) = \frac{1}{m}\sum_{i=1}^mg_{i,w}^\text{\tiny{CS}}(\boldsymbol\theta_t)$, and subtract it from the output of the CS algorithm, $\gv_w^{m.r.}(\boldsymbol\theta_t) = \gv_w^\text{\tiny{CS}}(\boldsymbol\theta_t) - \mu_w(t)\mathbf{1}_m$.
Then the channel input takes the form,
\begin{align}
    \xv_w(t) = \Big[ \sqrt{a_w(t)} \ \ \sqrt{a_w(t)}\mu_w(t) \ \ \sqrt{a_w(t)} \gv_w^{m.r.}(\boldsymbol\theta_t)\T \Big]\T ,
    \label{eq:chanIn}
\end{align}
where $a_w(t)$ is a scale factor chosen to satisfy the average symbol power constraint.
\begin{equation}
\begin{split}
   \frac{\Vert \xv_w(t) \Vert_2^2}{m_t+2} &=
    a_w(t)\Big( 1 + \mu_w^2(t) + \Vert  \gv_w^{m.r.}(\boldsymbol\theta_t) \Vert_2^2\Big)
    = P .
\end{split}
\end{equation}
As a result, $a_w(t)$ is specified by
\begin{align}
    a_w(t) = \frac{P(m_t+2)}{1 + (m_t-1)\mu_w^2(t) +\Vert  \gv_w^\text{\tiny{CS}}(\boldsymbol\theta_t) \Vert_2^2} .
\end{align}
Finally, workers form $\xv_w(t)$ and transmit it over the MAC.

\subsection{Parameter Server}
\label{subsec:PS}

Because all workers send their data synchronously, the PS receives the sum of the $W$ transmitted signals.
The goal is to recover a $K$-sparse vector from observation vector $\yv(t)$.
We denote  the part of the received signal that corresponds to the sum of the measurement vectors  by  $\yv_3^{M
_t}(t) = [ y_3(t) \ \ y_4(t) \ \ \cdots \ \ y_{M_t}(t) ]$.
The first step that the PS takes is to add the second value of the channel output, i.e., $y_2(t) = \sum_{w=1}^W\sqrt{a_w(t)}\mu_w + z_2(t)$, to $\yv_3^{M_t}(t)$ and divide the result by $y_1(t) = \sqrt{a_w(t)} + z_1(t)$.
The resulting vector is given by
\begin{align}
    \tilde{\yv}(t) = \frac{1}{y_1(t)}\big(\yv_3^{M_t}(t) + y_{2}(t)\mathbf{1}_{{M_t}-3}\big) . \label{eq:preProce}
\end{align}
Then, $\tilde{\yv}(t)$ is passed to the recovery algorithm.
The objective is to output a $K$-sparse vector, which hopefully contains the largest entries of the sum of the gradients.
Recall that the sum of the $K$-sparse vectors is likely to be a vector with $K^{'} \in [K:KW]$ non-zero entries.
As a result, $K^{'} - K$ entries of $\gv^{\text{\tiny{K}}}(\boldsymbol\theta_t)$ act as noise.
Let us define $\operatorname{supp}\big(\gv^{\text{\tiny{K}}}(\boldsymbol\theta_t)\big)$ and ${\cal A}$  to be the support of $\gv^{\text{\tiny{K}}}(\boldsymbol\theta_t)$ and the indices that correspond to the largest values of $\gv^{\text{\tiny{K}}}(\boldsymbol\theta_t)$, i.e.,
\begin{align*}
    &\operatorname{supp} \big(\gv^{\text{\tiny{K}}}(\boldsymbol\theta_t)\big)  = \{j : g_j^{\text{\tiny{K}}}(t) \neq 0\} \text{ and }
    {\cal A}  = \operatorname{sp}_K \big(\gv^{\text{\tiny{K}}}(\boldsymbol\theta_t)\big) .
\end{align*}
\begin{figure}
    \centering
 \scalebox{0.5}{\begin{tikzpicture}
 
Circle with label at 225 deg.
\node[draw,
    circle,
    minimum size =4cm,
    fill=red!50] (circle1) at (5,2.5){};
 
 \node at (5,5) {The set of all indices $[N] = \mathcal{A}\cup \mathcal{B}\cup \mathcal{F} $};

 \draw[-] (circle1.center) -- (circle1.south west);
  \draw[-] (circle1.center) -- (circle1.60);

\draw[fill=red!40] (5,2.5) -- +(270:2) arc (270:-90:2);
\draw[fill=violet!40] (5,2.5) -- +(235:2) arc (235:-90:2);
\draw[fill=yellow!40] (5,2.5) -- +(60:2) arc (60:-90:2);
\node at (4.4,1.1) {$\mathcal{A}$};
\node at (4.35,3) {$\mathcal{B}$};
\node at (6,2) {$\mathcal{F}$};
 \draw[-] (circle1.center) -- (circle1.south);
\node at (10,3)    {$\text{supp}\big(\gv^{\text{\tiny{K}}}(\boldsymbol\theta_t)\big) = \mathcal{A} \cup \mathcal{B}$,};
\node at (10.2,2.5) {False Alarm Set $=\mathcal{B}\cup \mathcal{F}$};

\draw [->] (3.5,0.5) -- (4.2,1);
\node at (3,0.2)    {$\operatorname{sp}_K \big(\gv^{\text{\tiny{K}}}(\boldsymbol\theta_t)\big)$};

\draw [->] (7.3,0.5) -- (6.2,1.8);
\node at (7,0.2)    {$\operatorname{zeros} \big(\gv^{\text{\tiny{K}}}(\boldsymbol\theta_t)\big)$};

\end{tikzpicture}}
        \caption{Illustration of Active Indices in the False Alarm Set, where $\mathcal{F}$ is the set of the non-active indices of $\gv^{\text{\tiny{K}}}(\boldsymbol\theta_t)$.}
        \label{fg:activeFADiagram} 
\end{figure}
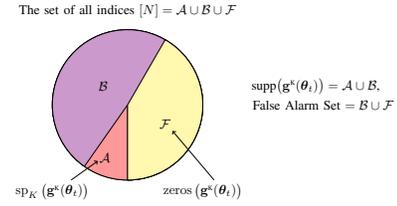
Then, the goal of the algorithm is to recover the indices in ${\cal A}$ and the corresponding values.
Nevertheless, if the decoder outputs an index that belongs to ${\cal B} = \operatorname{supp}\big(\gv^{\text{\tiny{K}}}(\boldsymbol\theta_t))\big) \backslash {\cal A}$ then, from training point of view, this index will contribute to the convergence of the CNN.
We define ${\cal B}$ as the \textit{Active Indices in the False Alarm Set}.
For example, let $\gv^{\text{\tiny{K}}}(\boldsymbol\theta_t)  = [-3, 1, 0, -0.01, 4, 0, 0, 0.5]$ and $K = 2$, then $\mathcal{A} = \{1,5\}$ and $\mathcal{B} = \{2,4,8\}$.
Figure~\ref{fg:activeFADiagram} shows the relation between the sets.
After this step, an estimate of the gradient $\gv^{\text{\tiny{K}}}(\boldsymbol\theta_t)$ is available and the PS updates the parameters as
$\boldsymbol{\theta}_{t+1} = \boldsymbol{\theta}_t - \alpha_t\hat{\gv}^{\text{\tiny{K}}}(\boldsymbol\theta_t)$,
where $\alpha_t$ is the learning rate and $\hat{\gv}^{\text{\tiny{K}}}(\boldsymbol\theta_t)$ denotes the estimate.
Finally, the updated model is broadcast to the workers, and the $t+1$ communication round begins.

\section{Compressed Sensing Scheme}
\label{sc:CSalgo}
In this section, we propose the CS algorithm that is used to compress the sparse gradient.
Throughout this section, the dependence of $t$ is omitted for notational convenience.
Since our scheme borrows ideas from \cite{gkagkosISIT2021}, we only outline the main functionality here.
We want to focus on the behaviour of the algorithm in the FL problem, and not on the structure of the CS algorithm itself.

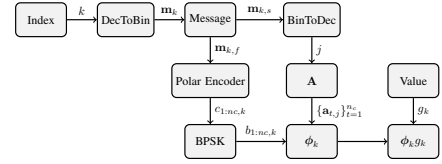
\begin{figure}
    \centering
 \scalebox{0.45}{\definecolor{darkgreen}{rgb}{0.01, 0.75, 0.24}
\definecolor{darkorange}{rgb}{0.77, 0.01, 0.2}
\definecolor{scarlet}{rgb}{1.0, 0.13, 0.0}

\begin{tikzpicture}
  [
  font=\normalsize, draw=black, line width=1pt,
  block/.style={rectangle, minimum height=10mm, minimum width=15mm,
  draw=black, fill=gray!10, rounded corners},
  entry/.style={rectangle, draw, inner sep=0pt, minimum size=2.5mm},
  ED/.style={rectangle, minimum height=10mm, minimum width=15mm, draw=black, rounded corners},
  symbol/.style={rectangle, draw, inner sep=0pt, minimum size=2.5mm},
  message/.style={rectangle, minimum height=5.5mm, minimum width=15mm, draw=black, rounded corners},
  symbol0/.style={rectangle, draw, fill=white, inner sep=0pt, minimum size=2.5mm},
    symbol1/.style={rectangle, draw, fill=blue!50, inner sep=0pt, minimum size=2.5mm},
    symbol2/.style={rectangle, draw, fill=white, inner sep=0pt, minimum size=2.5mm}
  ]

\node[block] (idx)  {Index};
\node[block,right of= idx, node distance = 2.5cm] (dectobin) {DecToBin};
\draw[->] (idx) -- (dectobin) node[midway,sloped,above] {$k$};
\node[block,right of= dectobin, node distance = 2.5cm] (msg)  {Message};
\draw[->] (dectobin) -- (msg) node[midway,sloped,above]{${\bf m}_k$};
\node[block,below of= msg, node distance = 1.8cm] (polar){Polar Encoder};
\node[block,right of= msg, node distance = 3.0cm] (bintodec) {BinToDec};

\draw[->] (msg) -- (polar) node[midway,sloped,right,rotate=90]{${\bf m}_{k,f}$};
\draw[->] (msg) -- (bintodec) node[midway,sloped,above]{${\bf m}_{k,s}$};

\def\xoffset{1.25}
\def\yoffset{-4.75}
\node[block,below of= polar, node distance = 1.8cm] (bpsk){BPSK};
\draw[->] (polar) -- (bpsk)node[midway,sloped,right,rotate=90]{$ c_{1:nc,k}$};

\node[block,below of= bintodec, node distance = 1.8cm] (A){${\bf A}$};
\draw[->] (bintodec) -- (A) node[midway,sloped,right,rotate=90]{$j$};

\node[block,below of= A, node distance = 1.8cm] (phi){$\boldsymbol\phi_k$};
\draw[->] (A) -- (phi)node[midway,sloped,right,rotate=90]{$\{{\bf a}_{t,j}\}_{t=1}^{n_c}$};
\draw[->] (bpsk) -- (phi)node[midway,sloped,above]{$b_{1:nc,k}$};

\node[block,right of= A, node distance = 3.0cm] (value){Value};
\node[block,below of= value, node distance = 1.8cm] (output){$\boldsymbol\phi_k g_k$};
\draw[->] (value) -- (output)node[midway,sloped,right,rotate=90]{$g_k$};
\draw[->] (phi) -- (output);

\end{tikzpicture}}
        \caption{Encoding procedure of the $k$th active index}
        \label{fg:encodeK} 
\end{figure}

\subsection{Compression at the Workers Side}
The model in the CS literature has the form $\yv = \boldsymbol\Phi\xv + \nv$, where $\xv \in \mathbb{R}^N$ is a $K$-sparse vector, $\boldsymbol{\Phi} \in \mathbb{R}^{M \times N}$ is a short-wide  matrix, $\nv$ is the measurement noise, and $\yv$ is the measurement vector.
The designer of a  CS algorithm aims to construct a measurement matrix and a recovery algorithm to solve a particular CS problem.
In our scheme, one can generate the columns of the measurement matrix $\boldsymbol\Phi$ on the fly.
Note that the active indices are available after the sparsification step.
Hence, each worker only needs to generate the active columns of $\boldsymbol{\Phi}$, thereby avoiding the vector-matrix multiplication.
Along these lines, we describe how one can generate the $k$th column of $\boldsymbol{\Phi}$ using the binary representation of $k$.
The encoding process of the $k$th index is illustrated in Fig.~\ref{fg:encodeK}.

Every vector $\gv_w^{\text{\tiny{K}}}(\boldsymbol\theta)$ has $K$ non-zero entries and, hence, it can be constructed as the weighted sum of $K$ standard basis vectors.
For each such vector summand, the integer location $k$ of the non-zero entry admits a binary representation $\mv_k$.
The length of the the binary sequence $\mv_k$ is $B = \lceil \log_2 N \rceil$ bits, possibly with leading zeros.
Every such binary sequence is split into two parts: $\mv_{k,f}$ and $\mv_{k,s}$ of lengths $B_f$ and $B_s$, respectively.
Based on the decimal representation of $\mv_{k,s}$, index~$k$ chooses one of the $J$
columns within the spreading dictionary $\mathbf{A}_i \in \{\pm \sqrt{1/N}\}^{L \times J}$, $i \in [n_c]$,
where $n_c$ represents the length of the code and $L$
is the length of the spreading sequences.
The entries of $\mathbf{A}_i$ are drawn independently from the set $\{\pm \sqrt{1/N}\}$ with equal probability.
The actual spreading operation for the $i$th coded bit is described next.
The decimal representation of $\mv_{k,s}$ is also employed to pick the values of the frozen bits for polar encoding.

To facilitate list decoding,  $\mv_{k,f}$ is padded with $r$ cyclic redundancy check (CRC) bits resulting in a message length of $r+B_f$ bits. 
A polar encoder maps this CRC augmented sequence into a codeword $\cv_{k}$ of length $n_c$.
Each coded bit of $c_{i,k}$, $i \in [n_c]$ is then BPSK modulated, $b_{i,k} \in \{ -1,+1\}$.
Finally, $b_{i,k}$ is spread by the spreading sequence $\av_{i,k}$, which is determined by $\mv_{k,s}$.
Given this encoding structure, we can see that the measurement matrix $\mathbf{\Phi}$ is composed of columns of the form,
\begin{align*}
\begin{bmatrix}
b_1 \av_{1,j}\T & b_2 \av_{2,j}\T &
\cdots & b_{n_c} \av_{n_c,j}\T
\end{bmatrix}\T
\end{align*}
where $j \in [J]$.
Next, the worker can apply the same methodology to generate the next active column.
After $K$ iterations, the worker has generated the active columns, and it can therefore obtain the compressed version of $\gv_w^{\text{\tiny{K}}}(\boldsymbol\theta)$ as
\begin{align}
   \gv_w^\text{\tiny{CS}}(\boldsymbol\theta) = \sum_{k \in {\cal K}_w}\boldsymbol\phi_k g_{k,w}^{\text{\tiny{K}}}(\boldsymbol\theta)
   =\boldsymbol\Phi \gv_w^{\text{\tiny{K}}}(\boldsymbol\theta) ,
\end{align}
where ${\cal K}_w$ is the support of $\gv_w^{\text{\tiny{K}}}(\boldsymbol\theta)$.
Since all the workers apply the same encoding strategy, the sum of $W$ such vectors is given by
\vspace{-5mm}
\begin{align*}
    \gv^\text{\tiny\text{\tiny{CS}}}(\boldsymbol\theta)
    &= \sum_{w=1}^W\boldsymbol\Phi \gv_w^{\text{\tiny{K}}}(\boldsymbol\theta) = \boldsymbol\Phi\Big(\sum_{w=1}^W \gv_w^{\text{\tiny{K}}}(\boldsymbol\theta)\Big) = \boldsymbol\Phi \gv^{\text{\tiny{K}}}(\boldsymbol\theta)\\
    & =  \underbrace{ \boldsymbol\Phi \gv_\mathcal{A}(\boldsymbol\theta)}_{\hbox{top-$K$ entries}} + \ \underbrace{ \boldsymbol\Phi \gv_\mathcal{B}(\boldsymbol\theta),}_{\hbox{other non-zero}}
\end{align*}
where $\gv_\mathcal{A}(\boldsymbol\theta)$ is the vector that the PS aims to recover.

\vspace{-3mm}
\subsection{Recovery Algorithm at the PS}
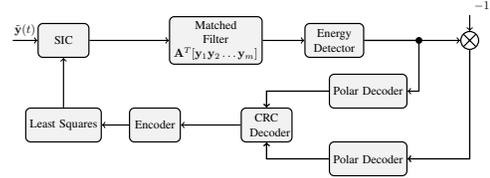
\begin{figure}[ht]
    \centering
 \scalebox{0.45}{\definecolor{myColor}{RGB}{36, 173, 71}
\begin{tikzpicture}
  [
  font=\normalsize, draw=black, line width=1pt,
  block/.style={rectangle, minimum height=10mm, minimum width=15mm,
  draw=black, fill=gray!10, rounded corners},
  sum/.style={rectangle, minimum height=15mm, minimum width=15mm, draw=black, rounded corners, fill=gray!10},
  message/.style={rectangle, minimum height=5.5mm, minimum width=80mm, draw=black, rounded corners},
  submessage/.style={rectangle, minimum height=5.5mm, minimum width=40mm, draw=black, rounded corners},
  inputCh/.style={rectangle, minimum height=5.5mm, minimum width=20mm, draw=black, rounded corners}
  ]
  \tikzstyle{product}   = [circle, minimum width=15pt, draw, inner sep=0pt, path picture={\draw (path picture bounding box.north west) -- (path picture bounding box.south east) (path picture bounding box.south west) -- (path picture bounding box.north east);}]

\node[block,align=center] (sic)  {SIC};
\draw[->] (-1.5,0) -- (sic.west) node[midway,above]{$\tilde{\yv}(t)$};
\node[block,right of =sic,node distance = 4.5cm,text width=2.5cm,align=center] (mf) {Matched \\  Filter \\ $\Am^T [\yv_1 \yv_2 \ldots \yv_m]$};

\draw[->] (sic) -- (mf);

\node[block,right of =mf,node distance = 3.5cm,text width=1.5cm,align=center] (ed) {Energy \\  Detector};
\draw[->] (mf) -- (ed);

\node[circle,draw, right of=ed,text=white,fill=black, scale=0.5,node distance = 2.5cm] (c1) {};
\draw[-] (ed) -- (c1) ;

\draw[-] (ed) -- (c1);

 \node [product, right of=c1, node distance = 1.5cm] (prod) {};
 \draw[->] (c1) -- (prod);
  \draw[->] (12,0.75) -- (prod.north) node[midway,right,yshift=0.5cm]{$-1$};
 \node[block,below of=ed, node distance = 1.5cm,xshift = 1cm] (polar1)  {Polar Decoder};
\node[block,below of=polar1, node distance = 2cm] (polar2)  {Polar Decoder};
 \draw[->] (c1) |- (polar1);
   \draw[->] (prod) |- (polar2);

  \draw[->] (prod) |- (polar2);

 \node [block,  below of = mf, node distance = 2.5cm,xshift = 1.5cm,text width=1cm,align=center] (crc) {CRC\\ Decoder};
 \draw[->] (polar1) -| (crc);
 \draw[->] (polar2) -| (crc);

\draw[->] (polar1) -| (crc);
\draw[->] (polar2) -| (crc);

\node [block,  left of = crc, node distance = 3.3cm] (en) {Encoder};
\draw[->] (crc) -- (en);


\node [block,  below of =sic, node distance = 2.5cm] (ls) {Least Squares};
\draw[->] (en) -- (ls) ;
\draw[->] (ls) -- (sic);




\end{tikzpicture}}
        \caption{The block diagram of the Recovery Algorithm.}
        \label{fg:recovryAlg} 
\end{figure}

The PS receives $\yv$ found in \eqref{eq:RXsignal} and, after the pre-processing step discussed in Section~\ref{subsec:PS}, the input to the recovery algorithm is given by \eqref{eq:preProce}.
The iterative recovery algorithm features a few distinct components; a matched filter, an energy detector, two polar decoders, a least squares estimator, and a successive interference canceller (SIC).
We should stress, again, that the sum of $K$-sparse signals need not be $K$-sparse.
However, the objective of the algorithm is to recover the largest $K$ entries of $\gv^\text{\tiny\text{\tiny{K}}}(\boldsymbol\theta)$ or, equivalently, ${\cal A}$.
A block diagram of the recovery algorithm is shown in Fig.~\ref{fg:recovryAlg}.

\subsubsection{Matched Filter and Energy Detector}
The signal $\tilde{\yv}$ is reshaped in a form amenable to sequence detection as follows,
\begin{align}
    \mathbf{\tilde{Y}} =
    \begin{bmatrix}
    \tilde{\yv}_1 & \tilde{\yv}_2 & \dots & \tilde{\yv}_{n_c}
    \end{bmatrix}
\end{align}
where $\tilde{\yv}_i \in \mathbb{R}^L$.
Section $\tilde{\yv}_i$ denotes the received signal corresponding to the $i$th polar coded symbol.
The sequence detector computes $ Z_{i,j} = \langle\av_{i,j} \tilde{\yv}_i \rangle$
for every pair $(i,j) \in [n_c] \times [J]$, where $\av_{i,j}$ is the $j$th spreading sequence from the set $\mathbf{A}_i$.
We call $Z_{i,j}$ the estimate of the $i$th coded bit, given that $\av_{i,j}$ is active.
The sequence detector iterates over all $(i,j)$ pairs and picks $K - S$ sequence with the largest $E_j = \sum_{i=1}^{n_c}|Z_{i,j}|^2$, where $S$ is the number of indices that the iterative algorithm has subtracted in the previous SIC iterations.

\subsubsection{Detection of Polar Codewords}

When spreading sequence $\av_{:,j}$ is active, the elements of $Z_{1:n_c,j}$ can act as estimates for the polar coded bits $b_{1:n_c,j}$.
Yet, the sign of $x_k$ is unknown and, hence, there is a need to run two list decoders.
The inputs to these list decoders are $Z_{1:n_c,j}$ and $-Z_{1:n_c,j}$, respectively.
The list decoder verifies the CRC constraint for every decoded codeword.
If two or more messages satisfy the checks, the most likely message is passed to the next step.

\subsubsection{Estimation of Non-Zero Entries in $\gv^{\text{\tiny{CS}}}(\boldsymbol\theta)$}
Since no prior distribution is available for the values of the gradient and $K \ll N$, we can apply least-squares (LS) estimation.

\subsubsection{Successive Interference Cancellation}
The contributions of the recovered non-zero entries in the sparse signal are removed from the received signal in the spirit of SIC.
The residual is then passed to the energy detector for the next decoding round.
This process continues until $K$ indices are recovered successfully, or there is no improvement between consecutive rounds, or the maximum number of iterations is reached.

\vspace{-5mm}

\vspace{-5mm}
\section{Experiments}
\label{sc:expe}
To evaluate the performance of the algorithm, we use it to solve a classification problem.
Specifically, the PS and workers wish to train a CNN to classify the images of the CIFAR-10 dataset~\cite{cifar}.
This dataset contains 60,000 RGB images of size $32 \times 32$ pixels, divided into 10 classes.
Since the goal of this article is to design a FL scheme for a large model, we use the ResNet20 \cite{ResNet20} to classify these images.
This network consists of $N=269,722$ parameters; we select $K=270$.
The FL setting includes a PS, $W=8$ workers and, every worker possesses $|\mathbf{D}_w| = 6144$ examples.
The data samples are chosen randomly and are distinct at each worker (i.i.d.\ datasets)\footnote{Since the compression algorithm does not exploit the correlation between the gradients at the worker side, it can be applied to FL problems with non-i.i.d.\ datasets. We anticipate a decrease in performance in such cases.}.
We use $|\mathbf{D}_{\text{train}}| = W|\mathbf{D}_w| = 49,152$ examples for the training, and $|\mathbf{D}_{\text{test}}| = 10,000$ for testing.
The batch size is $B_w(t) = 256$, and the number of communication rounds per epoch is $C = 24$.
The parameters are updated by the ADAM optimizer \cite{kingma2017adam} with learning rate $\text{lr} = 0.01$.
We compare our algorithm with a scenario, where a genie gives the average of the $W$ sparse vectors to the PS, i.e., $\frac{1}{W}\gv^{\text{\tiny{K}}}(\boldsymbol\theta_t)$ is directly available at the PS (no compression).
The PS uses $\frac{1}{W}\gv^{\text{\tiny{K}}}(\boldsymbol\theta_t)$ to update the model.
We denote this scheme by \textit{Dense}.
We set the target test accuracy to be $0.8$, and we count the number of channel uses needed from each scheme to achieve the target accuracy.

For PolarAir, we choose the initial parameters to be $B_f = 10$, $B_s = 9$, $J = 2^{B_f} = 1024$, $L = 400$, $n_c = 32$, $n_L = 2$ and $P=1000$.
Therefore, the initial number of measurements is $m_1 = 12,800$, which is 20 time less than $N$.
Let ${\cal \hat{A}}_{c,e}$ be the set of recovered indices after communication round $c \in [C]$ and during epoch $e$.
We increase the number of measurements if, at the end of the epoch, $Q \triangleq \frac{1}{C}\sum_{c=1}^C|{\cal \hat{A}}_{c,e}| \leq \frac{K}{2}$.
A naive policy to increase the number of measurements, $M = Ln_c$, is given by
\begin{align*}
    (L,n_c) = \begin{cases}
    (L,n_c + 32), \ \ & \text{if $Q \leq \frac{K}{2}$ for the first time}\\
    (L +100,n_c), \ \ & \text{otherwise} .
    \end{cases}
\end{align*}
That means, that the first time $Q \leq \frac{K}{2}$, PolarAir increases the length of the codeword $n_c$.
Then, $n_c$ remains fixed, and only the length of the spreading sequence grows.
A more optimal way to design $Q$ and the update policy is left for future work.

Figure~\ref{fg:testCU} depicts the test accuracy as a function of the channel uses.
The figure clearly shows that the CS based scheme outperforms \textit{Dense}.
On the other hand, Fig.~\ref{fg:testEpoc} presents the test performance of the two schemes as a function of the number of epochs.
As we expected, \textit{Dense} converges faster than PolarAir.
From a wireless communication point of view, we are interested in the number of channel uses rather than the number of epochs; this better captures the radio resources (spectrum and energy) utilized by the workers.
It should be noted that we do not compare our scheme with the one in \cite{Amiri-AMP-AWGN} because it is impractical to store a Gaussian measurement matrix with dimensions $M \times N$, where $N = 269,722$ and $M \geq 12800$; their algorithm cannot be implemented for the present setting.
We give several graphics in the rest of this section that illustrate the behavior of the proposed scheme and the sum of the sparse vectors during training.
These findings will aid in the development of task-specific compression algorithms.

    \begin{figure}
    \centering
    \begin{minipage}[b]{0.45\linewidth}
 \scalebox{0.45}{\begin{tikzpicture}
\definecolor{mycolor1}{rgb}{0.63529,0.07843,0.18431}%
\definecolor{mycolor2}{rgb}{0.00000,0.44706,0.74118}%
\definecolor{mycolor3}{rgb}{0.00000,0.49804,0.00000}%
\definecolor{mycolor4}{rgb}{0.87059,0.49020,0.00000}%
\definecolor{mycolor5}{rgb}{0.00000,0.44700,0.74100}%
\definecolor{mycolor6}{rgb}{0.74902,0.00000,0.74902}%
\definecolor{mycolor7}{rgb}{0.502,0.2000,0.5902}

\begin{axis}[
font=\footnotesize,
width=7cm,
height=5.5cm,
scale only axis,
xmin=0.5e6,
xmax=8e6,
xtick = {1e6,2e6,...,8e6},
xlabel={Channel Uses $(\times 24)$},
xmajorgrids,
ymin=0.1,
ymax=0.9,
ytick = {0.1,0.2,...,0.9},
ylabel={Test Accuracy},
ylabel near ticks,
ymajorgrids,
legend style={font=\scriptsize, at={(1,0.5)},anchor=north east, draw=black,fill=white,legend cell align=left}
]

\addplot [color=mycolor3,solid,line width=1pt]
  table[row sep=crcr]{
12800 0.8\\
8091660 0.8\\
  };
\addlegendentry{Target Test Acc. = 0.8};

\addplot [color=mycolor1,solid,line width=1pt]
  table[row sep=crcr]{
269722	3.13E-01	\\
539444	3.91E-01	\\
809166	4.35E-01	\\
1078888	4.94E-01	\\
1348610	5.38E-01	\\
1618332	5.76E-01	\\
1888054	6.01E-01	\\
2157776	6.30E-01	\\
2427498	6.50E-01	\\
2697220	6.74E-01	\\
2966942	6.86E-01	\\
3236664	7.03E-01	\\
3506386	7.18E-01	\\
3776108	7.27E-01	\\
4045830	7.45E-01	\\
4315552	7.46E-01	\\
4585274	7.54E-01	\\
4854996	7.67E-01	\\
5124718	7.78E-01	\\
5394440	7.73E-01	\\
5664162	7.88E-01	\\
5933884	7.83E-01	\\
6203606	7.87E-01	\\
6473328	8.01E-01	\\
6743050	8.04E-01	\\
7012772	8.00E-01	\\
7282494	8.03E-01	\\
7552216	8.08E-01	\\
7821938	8.12E-01	\\
8091660	8.14E-01	\\
};
\addlegendentry{Dense $K = 269,722$};

\addplot [color=mycolor2,solid,line width=1pt]
  table[row sep=crcr]{
12800	1.68E-01	\\
25600	2.28E-01	\\
38400	2.56E-01	\\
51200	2.85E-01	\\
64000	2.96E-01	\\
76800	3.09E-01	\\
89600	3.19E-01	\\
102400	3.34E-01	\\
115200	3.42E-01	\\
128000	3.52E-01	\\
140800	3.67E-01	\\
153600	3.72E-01	\\
166400	3.80E-01	\\
179200	3.98E-01	\\
192000	3.86E-01	\\
204800	4.04E-01	\\
217600	4.05E-01	\\
230400	4.15E-01	\\
243200	4.30E-01	\\
256000	4.28E-01	\\
268800	4.35E-01	\\
281600	4.43E-01	\\
294400	4.50E-01	\\
307200	4.49E-01	\\
320000	4.66E-01	\\
332800	4.58E-01	\\
345600	4.72E-01	\\
358400	4.76E-01	\\
371200	4.81E-01	\\
384000	4.86E-01	\\
396800	4.93E-01	\\
409600	4.97E-01	\\
422400	5.03E-01	\\
435200	5.10E-01	\\
448000	5.07E-01	\\
460800	5.18E-01	\\
473600	5.21E-01	\\
488000	5.19E-01	\\
502400	5.25E-01	\\
516800	5.28E-01	\\
531200	5.31E-01	\\
545600	5.35E-01	\\
560000	5.40E-01	\\
574400	5.42E-01	\\
588800	5.45E-01	\\
603200	5.57E-01	\\
619200	5.49E-01	\\
635200	5.56E-01	\\
651200	5.56E-01	\\
667200	5.60E-01	\\
683200	5.64E-01	\\
699200	5.67E-01	\\
715200	5.73E-01	\\
731200	5.75E-01	\\
747200	5.73E-01	\\
763200	5.77E-01	\\
779200	5.82E-01	\\
795200	5.92E-01	\\
811200	5.95E-01	\\
827200	5.94E-01	\\
843200	6.05E-01	\\
859200	6.05E-01	\\
875200	6.02E-01	\\
891200	6.05E-01	\\
907200	6.07E-01	\\
923200	6.12E-01	\\
939200	6.14E-01	\\
956800	6.11E-01	\\
974400	6.16E-01	\\
992000	6.14E-01	\\
1009600	6.11E-01	\\
1027200	6.23E-01	\\
1046400	6.22E-01	\\
1065600	6.23E-01	\\
1084800	6.25E-01	\\
1104000	6.22E-01	\\
1123200	6.30E-01	\\
1142400	6.30E-01	\\
1161600	6.32E-01	\\
1180800	6.36E-01	\\
1200000	6.32E-01	\\
1219200	6.45E-01	\\
1238400	6.47E-01	\\
1257600	6.52E-01	\\
1276800	6.49E-01	\\
1296000	6.52E-01	\\
1315200	6.57E-01	\\
1334400	6.52E-01	\\
1353600	6.59E-01	\\
1372800	6.62E-01	\\
1392000	6.58E-01	\\
1411200	6.62E-01	\\
1432000	6.65E-01	\\
1452800	6.60E-01	\\
1473600	6.65E-01	\\
1494400	6.69E-01	\\
1516800	6.71E-01	\\
1539200	6.75E-01	\\
1561600	6.77E-01	\\
1584000	6.75E-01	\\
1606400	6.71E-01	\\
1628800	6.75E-01	\\
1651200	6.78E-01	\\
1673600	6.78E-01	\\
1696000	6.83E-01	\\
1718400	6.83E-01	\\
1740800	6.85E-01	\\
1763200	6.89E-01	\\
1785600	6.93E-01	\\
1808000	6.95E-01	\\
1830400	6.95E-01	\\
1852800	6.95E-01	\\
1875200	6.96E-01	\\
1897600	6.99E-01	\\
1920000	7.02E-01	\\
1942400	7.01E-01	\\
1964800	7.05E-01	\\
1987200	7.06E-01	\\
2009600	7.03E-01	\\
2032000	7.07E-01	\\
2054400	7.07E-01	\\
2076800	7.12E-01	\\
2099200	7.14E-01	\\
2121600	7.12E-01	\\
2144000	7.13E-01	\\
2166400	7.17E-01	\\
2188800	7.19E-01	\\
2211200	7.19E-01	\\
2233600	7.24E-01	\\
2256000	7.23E-01	\\
2278400	7.23E-01	\\
2300800	7.26E-01	\\
2323200	7.29E-01	\\
2345600	7.28E-01	\\
2368000	7.28E-01	\\
2392000	7.29E-01	\\
2416000	7.28E-01	\\
2440000	7.29E-01	\\
2464000	7.30E-01	\\
2488000	7.30E-01	\\
2512000	7.32E-01	\\
2536000	7.36E-01	\\
2560000	7.39E-01	\\
2584000	7.36E-01	\\
2608000	7.40E-01	\\
2632000	7.40E-01	\\
2656000	7.39E-01	\\
2680000	7.44E-01	\\
2704000	7.46E-01	\\
2728000	7.43E-01	\\
2752000	7.45E-01	\\
2776000	7.49E-01	\\
2800000	7.44E-01	\\
2824000	7.47E-01	\\
2848000	7.46E-01	\\
2872000	7.51E-01	\\
2896000	7.52E-01	\\
2920000	7.52E-01	\\
2944000	7.55E-01	\\
2968000	7.56E-01	\\
2992000	7.57E-01	\\
3016000	7.56E-01	\\
3040000	7.53E-01	\\
3064000	7.59E-01	\\
3088000	7.61E-01	\\
3112000	7.56E-01	\\
3136000	7.56E-01	\\
3160000	7.57E-01	\\
3184000	7.63E-01	\\
3208000	7.64E-01	\\
3232000	7.62E-01	\\
3256000	7.62E-01	\\
3280000	7.63E-01	\\
3304000	7.59E-01	\\
3328000	7.64E-01	\\
3352000	7.64E-01	\\
3376000	7.66E-01	\\
3400000	7.65E-01	\\
3424000	7.66E-01	\\
3448000	7.67E-01	\\
3472000	7.68E-01	\\
3496000	7.67E-01	\\
3520000	7.71E-01	\\
3544000	7.73E-01	\\
3568000	7.72E-01	\\
3592000	7.72E-01	\\
3616000	7.74E-01	\\
3640000	7.74E-01	\\
3664000	7.75E-01	\\
3688000	7.76E-01	\\
3712000	7.75E-01	\\
3736000	7.75E-01	\\
3760000	7.75E-01	\\
3784000	7.76E-01	\\
3808000	7.80E-01	\\
3832000	7.78E-01	\\
3856000	7.79E-01	\\
3880000	7.78E-01	\\
3904000	7.79E-01	\\
3929600	7.79E-01	\\
3955200	7.77E-01	\\
3980800	7.76E-01	\\
4006400	7.78E-01	\\
4032000	7.76E-01	\\
4057600	7.78E-01	\\
4083200	7.80E-01	\\
4108800	7.82E-01	\\
4134400	7.80E-01	\\
4160000	7.82E-01	\\
4185600	7.84E-01	\\
4211200	7.84E-01	\\
4236800	7.85E-01	\\
4262400	7.84E-01	\\
4288000	7.84E-01	\\
4313600	7.85E-01	\\
4339200	7.86E-01	\\
4364800	7.90E-01	\\
4390400	7.88E-01	\\
4416000	7.89E-01	\\
4441600	7.87E-01	\\
4467200	7.85E-01	\\
4492800	7.90E-01	\\
4518400	7.92E-01	\\
4544000	7.90E-01	\\
4569600	7.87E-01	\\
4595200	7.90E-01	\\
4620800	7.92E-01	\\
4646400	7.91E-01	\\
4672000	7.91E-01	\\
4697600	7.92E-01	\\
4723200	7.94E-01	\\
4748800	7.93E-01	\\
4774400	7.95E-01	\\
4800000	7.94E-01	\\
4825600	7.96E-01	\\
4851200	7.96E-01	\\
4876800	7.97E-01	\\
4902400	7.99E-01	\\
4928000	7.94E-01	\\
4953600	7.97E-01	\\
4979200	7.95E-01	\\
5004800	7.98E-01	\\
5030400	8.00E-01	\\
5056000	7.98E-01	\\
5081600	7.98E-01	\\
5107200	8.01E-01	\\
5132800	8.02E-01	\\
5158400	8.00E-01	\\
5184000	8.00E-01	\\
5209600	8.02E-01	\\
5235200	8.00E-01	\\
5260800	7.99E-01	\\
5286400	8.02E-01	\\
5312000	8.05E-01	\\
5337600	8.04E-01	\\
5363200	8.04E-01	\\
5388800	8.04E-01	\\
5414400	8.06E-01	\\
5440000	8.03E-01	\\
5465600	8.01E-01	\\
5491200	8.04E-01	\\
5516800	8.03E-01	\\
5542400	8.05E-01	\\
5568000	8.06E-01	\\
5593600	8.05E-01	\\
5619200	8.06E-01	\\
5644800	8.07E-01	\\
5670400	8.07E-01	\\
5696000	8.05E-01	\\
5721600	8.10E-01	\\
5747200	8.08E-01	\\
5772800	8.08E-01	\\
5798400	8.10E-01	\\
5824000	8.10E-01	\\
5849600	8.09E-01	\\
5875200	8.09E-01	\\
5900800	8.08E-01	\\
5926400	8.08E-01	\\
5952000	8.08E-01	\\
5977600	8.10E-01	\\
6003200	8.09E-01	\\
6028800	8.10E-01	\\
6054400	8.10E-01	\\
6080000	8.10E-01	\\
6105600	8.13E-01	\\
6131200	8.13E-01	\\
6156800	8.12E-01	\\
6182400	8.13E-01	\\
6208000	8.11E-01	\\
6233600	8.11E-01	\\
6259200	8.11E-01	\\
6284800	8.14E-01	\\
6310400	8.12E-01	\\
6336000	8.14E-01	\\
6361600	8.14E-01	\\
6387200	8.16E-01	\\
6412800	8.14E-01	\\
6438400	8.15E-01	\\
6464000	8.15E-01	\\
6489600	8.16E-01	\\
};
\addlegendentry{PolarAir};

\end{axis}

\end{tikzpicture}%
}
        \caption{Test Accuracy as a function of Channel Uses. }
        \label{fg:testCU} 
    \end{minipage}
    \begin{minipage}[b]{0.45\linewidth}
 \scalebox{0.45}{\begin{tikzpicture}
\definecolor{mycolor1}{rgb}{0.63529,0.07843,0.18431}%
\definecolor{mycolor2}{rgb}{0.00000,0.44706,0.74118}%
\definecolor{mycolor3}{rgb}{0.00000,0.49804,0.00000}%
\definecolor{mycolor4}{rgb}{0.87059,0.49020,0.00000}%
\definecolor{mycolor5}{rgb}{0.00000,0.44700,0.74100}%
\definecolor{mycolor6}{rgb}{0.74902,0.00000,0.74902}%
\definecolor{mycolor7}{rgb}{0.502,0.2000,0.5902}

\begin{axis}[
font=\footnotesize,
width=7cm,
height=5.5cm,
scale only axis,
xmin=1,
xmax=300,
xtick = {0,50,...,300},
xlabel={Epochs},
xmajorgrids,
ymin=0.1,
ymax=0.9,
ytick = {0.1,0.2,...,0.9},
ylabel={Test Accuracy},
ylabel near ticks,
ymajorgrids,
legend style={font=\scriptsize, at={(1,0.5)},anchor=north east, draw=black,fill=white,legend cell align=left}
]

\addplot [color=mycolor3,solid,line width=1pt]
  table[row sep=crcr]{
1 0.8\\
300 0.8\\
};
\addlegendentry{Target Test Acc. = 0.8};

\addplot [color=mycolor1,solid,line width=1pt]
  table[row sep=crcr]{
1	3.13E-01	\\
2	3.91E-01	\\
3	4.35E-01	\\
4	4.94E-01	\\
5	5.38E-01	\\
6	5.76E-01	\\
7	6.01E-01	\\
8	6.30E-01	\\
9	6.50E-01	\\
10	6.74E-01	\\
11	6.86E-01	\\
12	7.03E-01	\\
13	7.18E-01	\\
14	7.27E-01	\\
15	7.45E-01	\\
16	7.46E-01	\\
17	7.54E-01	\\
18	7.67E-01	\\
19	7.78E-01	\\
20	7.73E-01	\\
21	7.88E-01	\\
22	7.83E-01	\\
23	7.87E-01	\\
24	8.01E-01	\\
25	8.04E-01	\\
26	8.00E-01	\\
27	8.03E-01	\\
28	8.08E-01	\\
29	8.12E-01	\\
30	8.14E-01	\\
};
\addlegendentry{Dense $K = 269,722$};

\addplot [color=mycolor2,solid,line width=1pt]
  table[row sep=crcr]{
1.00E+00	1.68E-01	\\
2.00E+00	2.28E-01	\\
3.00E+00	2.56E-01	\\
4.00E+00	2.85E-01	\\
5.00E+00	2.96E-01	\\
6.00E+00	3.09E-01	\\
7.00E+00	3.19E-01	\\
8.00E+00	3.34E-01	\\
9.00E+00	3.42E-01	\\
1.00E+01	3.52E-01	\\
1.10E+01	3.67E-01	\\
1.20E+01	3.72E-01	\\
1.30E+01	3.80E-01	\\
1.40E+01	3.98E-01	\\
1.50E+01	3.86E-01	\\
1.60E+01	4.04E-01	\\
1.70E+01	4.05E-01	\\
1.80E+01	4.15E-01	\\
1.90E+01	4.30E-01	\\
2.00E+01	4.28E-01	\\
2.10E+01	4.35E-01	\\
2.20E+01	4.43E-01	\\
2.30E+01	4.50E-01	\\
2.40E+01	4.49E-01	\\
2.50E+01	4.66E-01	\\
2.60E+01	4.58E-01	\\
2.70E+01	4.72E-01	\\
2.80E+01	4.76E-01	\\
2.90E+01	4.81E-01	\\
3.00E+01	4.86E-01	\\
3.10E+01	4.93E-01	\\
3.20E+01	4.97E-01	\\
3.30E+01	5.03E-01	\\
3.40E+01	5.10E-01	\\
3.50E+01	5.07E-01	\\
3.60E+01	5.18E-01	\\
3.70E+01	5.21E-01	\\
3.80E+01	5.19E-01	\\
3.90E+01	5.25E-01	\\
4.00E+01	5.28E-01	\\
4.10E+01	5.31E-01	\\
4.20E+01	5.35E-01	\\
4.30E+01	5.40E-01	\\
4.40E+01	5.42E-01	\\
4.50E+01	5.45E-01	\\
4.60E+01	5.57E-01	\\
4.70E+01	5.49E-01	\\
4.80E+01	5.56E-01	\\
4.90E+01	5.56E-01	\\
5.00E+01	5.60E-01	\\
5.10E+01	5.64E-01	\\
5.20E+01	5.67E-01	\\
5.30E+01	5.73E-01	\\
5.40E+01	5.75E-01	\\
5.50E+01	5.73E-01	\\
5.60E+01	5.77E-01	\\
5.70E+01	5.82E-01	\\
5.80E+01	5.92E-01	\\
5.90E+01	5.95E-01	\\
6.00E+01	5.94E-01	\\
6.10E+01	6.05E-01	\\
6.20E+01	6.05E-01	\\
6.30E+01	6.02E-01	\\
6.40E+01	6.05E-01	\\
6.50E+01	6.07E-01	\\
6.60E+01	6.12E-01	\\
6.70E+01	6.14E-01	\\
6.80E+01	6.11E-01	\\
6.90E+01	6.16E-01	\\
7.00E+01	6.14E-01	\\
7.10E+01	6.11E-01	\\
7.20E+01	6.23E-01	\\
7.30E+01	6.22E-01	\\
7.40E+01	6.23E-01	\\
7.50E+01	6.25E-01	\\
7.60E+01	6.22E-01	\\
7.70E+01	6.30E-01	\\
7.80E+01	6.30E-01	\\
7.90E+01	6.32E-01	\\
8.00E+01	6.36E-01	\\
8.10E+01	6.32E-01	\\
8.20E+01	6.45E-01	\\
8.30E+01	6.47E-01	\\
8.40E+01	6.52E-01	\\
8.50E+01	6.49E-01	\\
8.60E+01	6.52E-01	\\
8.70E+01	6.57E-01	\\
8.80E+01	6.52E-01	\\
8.90E+01	6.59E-01	\\
9.00E+01	6.62E-01	\\
9.10E+01	6.58E-01	\\
9.20E+01	6.62E-01	\\
9.30E+01	6.65E-01	\\
9.40E+01	6.60E-01	\\
9.50E+01	6.65E-01	\\
9.60E+01	6.69E-01	\\
9.70E+01	6.71E-01	\\
9.80E+01	6.75E-01	\\
9.90E+01	6.77E-01	\\
1.00E+02	6.75E-01	\\
1.01E+02	6.71E-01	\\
1.02E+02	6.75E-01	\\
1.03E+02	6.78E-01	\\
1.04E+02	6.78E-01	\\
1.05E+02	6.83E-01	\\
1.06E+02	6.83E-01	\\
1.07E+02	6.85E-01	\\
1.08E+02	6.89E-01	\\
1.09E+02	6.93E-01	\\
1.10E+02	6.95E-01	\\
1.11E+02	6.95E-01	\\
1.12E+02	6.95E-01	\\
1.13E+02	6.96E-01	\\
1.14E+02	6.99E-01	\\
1.15E+02	7.02E-01	\\
1.16E+02	7.01E-01	\\
1.17E+02	7.05E-01	\\
1.18E+02	7.06E-01	\\
1.19E+02	7.03E-01	\\
1.20E+02	7.07E-01	\\
1.21E+02	7.07E-01	\\
1.22E+02	7.12E-01	\\
1.23E+02	7.14E-01	\\
1.24E+02	7.12E-01	\\
1.25E+02	7.13E-01	\\
1.26E+02	7.17E-01	\\
1.27E+02	7.19E-01	\\
1.28E+02	7.19E-01	\\
1.29E+02	7.24E-01	\\
1.30E+02	7.23E-01	\\
1.31E+02	7.23E-01	\\
1.32E+02	7.26E-01	\\
1.33E+02	7.29E-01	\\
1.34E+02	7.28E-01	\\
1.35E+02	7.28E-01	\\
1.36E+02	7.29E-01	\\
1.37E+02	7.28E-01	\\
1.38E+02	7.29E-01	\\
1.39E+02	7.30E-01	\\
1.40E+02	7.30E-01	\\
1.41E+02	7.32E-01	\\
1.42E+02	7.36E-01	\\
1.43E+02	7.39E-01	\\
1.44E+02	7.36E-01	\\
1.45E+02	7.40E-01	\\
1.46E+02	7.40E-01	\\
1.47E+02	7.39E-01	\\
1.48E+02	7.44E-01	\\
1.49E+02	7.46E-01	\\
1.50E+02	7.43E-01	\\
1.51E+02	7.45E-01	\\
1.52E+02	7.49E-01	\\
1.53E+02	7.44E-01	\\
1.54E+02	7.47E-01	\\
1.55E+02	7.46E-01	\\
1.56E+02	7.51E-01	\\
1.57E+02	7.52E-01	\\
1.58E+02	7.52E-01	\\
1.59E+02	7.55E-01	\\
1.60E+02	7.56E-01	\\
1.61E+02	7.57E-01	\\
1.62E+02	7.56E-01	\\
1.63E+02	7.53E-01	\\
1.64E+02	7.59E-01	\\
1.65E+02	7.61E-01	\\
1.66E+02	7.56E-01	\\
1.67E+02	7.56E-01	\\
1.68E+02	7.57E-01	\\
1.69E+02	7.63E-01	\\
1.70E+02	7.64E-01	\\
1.71E+02	7.62E-01	\\
1.72E+02	7.62E-01	\\
1.73E+02	7.63E-01	\\
1.74E+02	7.59E-01	\\
1.75E+02	7.64E-01	\\
1.76E+02	7.64E-01	\\
1.77E+02	7.66E-01	\\
1.78E+02	7.65E-01	\\
1.79E+02	7.66E-01	\\
1.80E+02	7.67E-01	\\
1.81E+02	7.68E-01	\\
1.82E+02	7.67E-01	\\
1.83E+02	7.71E-01	\\
1.84E+02	7.73E-01	\\
1.85E+02	7.72E-01	\\
1.86E+02	7.72E-01	\\
1.87E+02	7.74E-01	\\
1.88E+02	7.74E-01	\\
1.89E+02	7.75E-01	\\
1.90E+02	7.76E-01	\\
1.91E+02	7.75E-01	\\
1.92E+02	7.75E-01	\\
1.93E+02	7.75E-01	\\
1.94E+02	7.76E-01	\\
1.95E+02	7.80E-01	\\
1.96E+02	7.78E-01	\\
1.97E+02	7.79E-01	\\
1.98E+02	7.78E-01	\\
1.99E+02	7.79E-01	\\
2.00E+02	7.79E-01	\\
2.01E+02	7.77E-01	\\
2.02E+02	7.76E-01	\\
2.03E+02	7.78E-01	\\
2.04E+02	7.76E-01	\\
2.05E+02	7.78E-01	\\
2.06E+02	7.80E-01	\\
2.07E+02	7.82E-01	\\
2.08E+02	7.80E-01	\\
2.09E+02	7.82E-01	\\
2.10E+02	7.84E-01	\\
2.11E+02	7.84E-01	\\
2.12E+02	7.85E-01	\\
2.13E+02	7.84E-01	\\
2.14E+02	7.84E-01	\\
2.15E+02	7.85E-01	\\
2.16E+02	7.86E-01	\\
2.17E+02	7.90E-01	\\
2.18E+02	7.88E-01	\\
2.19E+02	7.89E-01	\\
2.20E+02	7.87E-01	\\
2.21E+02	7.85E-01	\\
2.22E+02	7.90E-01	\\
2.23E+02	7.92E-01	\\
2.24E+02	7.90E-01	\\
2.25E+02	7.87E-01	\\
2.26E+02	7.90E-01	\\
2.27E+02	7.92E-01	\\
2.28E+02	7.91E-01	\\
2.29E+02	7.91E-01	\\
2.30E+02	7.92E-01	\\
2.31E+02	7.94E-01	\\
2.32E+02	7.93E-01	\\
2.33E+02	7.95E-01	\\
2.34E+02	7.94E-01	\\
2.35E+02	7.96E-01	\\
2.36E+02	7.96E-01	\\
2.37E+02	7.97E-01	\\
2.38E+02	7.99E-01	\\
2.39E+02	7.94E-01	\\
2.40E+02	7.97E-01	\\
2.41E+02	7.95E-01	\\
2.42E+02	7.98E-01	\\
2.43E+02	8.00E-01	\\
2.44E+02	7.98E-01	\\
2.45E+02	7.98E-01	\\
2.46E+02	8.01E-01	\\
2.47E+02	8.02E-01	\\
2.48E+02	8.00E-01	\\
2.49E+02	8.00E-01	\\
2.50E+02	8.02E-01	\\
2.51E+02	8.00E-01	\\
2.52E+02	7.99E-01	\\
2.53E+02	8.02E-01	\\
2.54E+02	8.05E-01	\\
2.55E+02	8.04E-01	\\
2.56E+02	8.04E-01	\\
2.57E+02	8.04E-01	\\
2.58E+02	8.06E-01	\\
2.59E+02	8.03E-01	\\
2.60E+02	8.01E-01	\\
2.61E+02	8.04E-01	\\
2.62E+02	8.03E-01	\\
2.63E+02	8.05E-01	\\
2.64E+02	8.06E-01	\\
2.65E+02	8.05E-01	\\
2.66E+02	8.06E-01	\\
2.67E+02	8.07E-01	\\
2.68E+02	8.07E-01	\\
2.69E+02	8.05E-01	\\
2.70E+02	8.10E-01	\\
2.71E+02	8.08E-01	\\
2.72E+02	8.08E-01	\\
2.73E+02	8.10E-01	\\
2.74E+02	8.10E-01	\\
2.75E+02	8.09E-01	\\
2.76E+02	8.09E-01	\\
2.77E+02	8.08E-01	\\
2.78E+02	8.08E-01	\\
2.79E+02	8.08E-01	\\
2.80E+02	8.10E-01	\\
2.81E+02	8.09E-01	\\
2.82E+02	8.10E-01	\\
2.83E+02	8.10E-01	\\
2.84E+02	8.10E-01	\\
2.85E+02	8.13E-01	\\
2.86E+02	8.13E-01	\\
2.87E+02	8.12E-01	\\
2.88E+02	8.13E-01	\\
2.89E+02	8.11E-01	\\
2.90E+02	8.11E-01	\\
2.91E+02	8.11E-01	\\
2.92E+02	8.14E-01	\\
2.93E+02	8.12E-01	\\
2.94E+02	8.14E-01	\\
2.95E+02	8.14E-01	\\
2.96E+02	8.16E-01	\\
2.97E+02	8.14E-01	\\
2.98E+02	8.15E-01	\\
2.99E+02	8.15E-01	\\
3.00E+02	8.16E-01	\\
};
\addlegendentry{PolarAir};
\end{axis}

\end{tikzpicture}%
}
        \caption{Test Accuracy as a function of Epochs.}
        \label{fg:testEpoc} 
    \end{minipage}
\end{figure}

First we investigate  the number of non-zero values in the sum of the gradients $\gv^K(\boldsymbol\theta_t)$.
Figure~\ref{fg:nonZeroEpoc} plots the average of the total number of active indices, i.e. $\mathcal{A} \cup \mathcal{B}$, (see Fig~\ref{fg:activeFADiagram}) as a function of epochs.
Note that $K=270$ and $W = 8$, hence there can be at most $KW = 2160$ non-zero values in $\gv^K(\boldsymbol\theta_t)$.
It is clear that after the $50$th epoch the average number of nonzero values is around $1900$.
Since the objective of the CS component is to retrieve $K=270$, the remaining $1900-K = 1630$ values essentially behave like noise.
This indicates that, as the number of epochs increases, so does the noise in the CS problem, creating the need for further measurements.

Next, we are interested in the performance of the CS scheme.
The Detection $\pd$ and False Alarm $\pfa$ probability of error of the recovery algorithm are shown in
Fig.~\ref{fg:DE} and Fig.~\ref{fg:FA}.
Interestingly, as the number of epochs increases, the   $\pd$  and $\pfa$  converge to 0.5 and 0.4, respectively.
Values that show poor performance from CS perspective.
However, the test accuracy of the model increases, (see Figures~\ref{fg:testCU} \&~\ref{fg:testEpoc}).
The reason behind this contradiction is given in Fig.~\ref{fg:actiFA}, where the average number of recovered indices that belongs to  \textit{Active Indices in the False Alarm Set}, i.e. $\hat{\mathcal{B}}$,  as a function of epochs is illustrated.
We denote  $\hat{\mathcal{B}}$ all the indices that the  recovery algorithm recovers and belongs to  $\mathcal{B}$.
Clearly, when this number increases, the performance of the recovery algorithm suffers.
However, because at least some of the workers communicate these indices, recovering them and using them for a gradient update is advantageous from a training standpoint.
Further analysis and additional comments are omitted due to space limitations.

\begin{figure}
    \centering
    \begin{minipage}[b]{0.45\linewidth}
 \scalebox{0.45}{\begin{tikzpicture}
\definecolor{mycolor1}{rgb}{0.63529,0.07843,0.18431}%
\definecolor{mycolor2}{rgb}{0.00000,0.44706,0.74118}%
\definecolor{mycolor3}{rgb}{0.00000,0.49804,0.00000}%
\definecolor{mycolor4}{rgb}{0.87059,0.49020,0.00000}%
\definecolor{mycolor5}{rgb}{0.00000,0.44700,0.74100}%
\definecolor{mycolor6}{rgb}{0.74902,0.00000,0.74902}%
\definecolor{mycolor7}{rgb}{0.502,0.2000,0.5902}

\begin{axis}[
font=\footnotesize,
width=7cm,
height=5.5cm,
scale only axis,
xmin=1,
xmax=300,
xtick = {0,50,...,300},
xlabel={Epochs},
xmajorgrids,
ymin=800,
ymax=2200,
ytick = {800,1000,1200,...,2200},
ylabel={Average Number of Active indices $(\mathcal{A} \cup \mathcal{B})$},
ylabel near ticks,
ymajorgrids,
legend style={font=\scriptsize, at={(1,1.4)},anchor=north east, draw=black,fill=white,legend cell align=left}
]

\addplot [color=mycolor1,solid,line width=1.5pt]
  table[row sep=crcr]{
1	1103.96875\\
2	1572.604167\\
3	1681.557292\\
4	1715.052083\\
5	1719.984375\\
6	1727.265625\\
7	1728.723958\\
8	1735.520833\\
9	1742.567708\\
10	1726.130208\\
11	1714.651042\\
12	1749.161458\\
13	1778.796875\\
14	1813.890625\\
15	1828.125\\
16	1832.14583\\
17	1836.041667\\
18	1841.005208\\
19	1840.822917\\
20	1862.244792\\
21	1848.041667\\
22	1860.989583\\
23	1859.067708\\
24	1864.822917\\
25	1868.291667\\
26	1855.3125\\
27	1856.427083\\
28	1859.338542\\
29	1882.223958\\
30	1905.083333\\
31	1902.015625\\
32	1902.005208\\
33	1907.552083\\
34	1915.286458\\
35	1924.71875\\
36	1915.760417\\
37	1933.505208\\
38	1906.375\\
39	1887.89062\\
40	1880.307292\\
41	1894.75\\
42	1900.1927\\
43	1912.609375\\
44	1911.84375\\
45	1920.822917\\
46	1918.322917\\
47	1902.494792\\
48	1879.177083\\
49	1886.78125\\
50	1874.005208\\
51	1873.109375\\
52	1870.546875\\
53	1886.598958\\
54	1896.411458\\
55	1905.302083\\
56	1897.59375\\
57	1901.677083\\
58	1918.791667\\
59	1928.03125\\
60	1942.260417\\
61	1943.234375\\
62	1942.895833\\
63	1924.838542\\
64	1931.765625\\
65	1943.84375\\
66	1942.46875\\
67	1948.447917\\
68	1933.098958\\
69	1907.890625\\
70	1908.541667\\
71	1902.338542\\
72	1909.90625\\
73	1913.40625\\
74	1882.104167\\
75	1876.1875\\
76	1874.25\\
77	1882.2812\\
78	1909.052083\\
79	1934.510417\\
80	1933.239583\\
81	1918.125\\
82	1927.34375\\
83	1933.052083\\
84	1923.864583\\
85	1906.958333\\
86	1892.6875\\
87	1886.25\\
88	1896.9322\\
89	1895.260417\\
90	1906.276042\\
91	1929.838542\\
92	1930.119792\\
93	1918.546875\\
94	1907.942708\\
95	1903.755208\\
96	1918\\
97	1927.1\\
98	1920.447\\
99	1900.98437\\
100	1895.369792\\
101	1891.364583\\
102	1907.1875\\
103	1911.307292\\
104	1884.333333\\
105	1900.838542\\
106	1903.557292\\
107	1904.625\\
108	1908.48958\\
109	1916.796875\\
110	1911.145833\\
111	1915.90625\\
112	1908.510417\\
113	1890.625\\
114	1887.21875\\
115	1888.255208\\
116	1893.53125\\
117	1900.947917\\
118	1904.947917\\
119	1903.09375\\
120	1897.286458\\
121	1898.661458\\
122	1898.083333\\
123	1912.494792\\
124	1918.807292\\
125	1902.546875\\
126	1909.59375\\
127	1926.916667\\
128	1907.703125\\
129	1908.348958\\
130	1914.036458\\
131	1919.239583\\
132	1932.953125\\
133	1928.973958\\
134	1930.526042\\
135	1926.135417\\
136	1910.984375\\
137	1912.421875\\
138	1897.65625\\
139	1890.692708\\
140	1881.786458\\
141	1892.817708\\
142	1908.770833\\
143	1906.541667\\
144	1893.614583\\
145	1888.083333\\
146	1910.21875\\
147	1908.84375\\
148	1895.989583\\
149	1906.083333\\
150	1910.020833\\
151	1909.473958\\
152	1917.640625\\
153	1917.203125\\
154	1931.875\\
155	1920.40625\\
156	1925.28125\\
157	1943.432292\\
158	1945.463542\\
159	1953.692708\\
160	1942.166667\\
161	1931.625\\
162	1936.56770\\
163	1910.838542\\
164	1906.598958\\
165	1909.828125\\
166	1902.661458\\
167	1907.197917\\
168	1897.5\\
169	1903.572\\
170	1912.88020\\
171	1908.614583\\
172	1908.380208\\
173	1911.328125\\
174	1912.364583\\
175	1914.338542\\
176	1911.583333\\
177	1910.432292\\
178	1926.984375\\
179	1915.734375\\
180	1918.421875\\
181	1923.088542\\
182	1921.192708\\
183	1912.989583\\
184	1927.182292\\
185	1917.5\\
186	1903.718\\
187	1915.78645\\
188	1925.390625\\
189	1921.848958\\
190	1932\\
191	1924.9\\
192	1924.369\\
193	1918.67187\\
194	1922.723958\\
195	1925.609375\\
196	1928.34375\\
197	1927.604167\\
198	1918.244792\\
199	1905.421875\\
200	1900.583333\\
201	1908.885417\\
202	1915.052083\\
203	1904.21875\\
204	1891.932292\\
205	1891.71875\\
206	1897.59375\\
207	1891.614583\\
208	1893.505208\\
209	1899.760417\\
210	1916.572917\\
211	1920.8125\\
212	1920\\
213	1920.2\\
214	1931.661\\
215	1930.08333\\
216	1942.25\\
217	1938.0312\\
218	1939.536458\\
219	1943.932292\\
220	1935.005208\\
221	1933.864583\\
222	1929.635417\\
223	1925.947917\\
224	1912.348958\\
225	1914.328125\\
226	1910\\
227	1907.6\\
228	1912.286\\
229	1911.97916\\
230	1921.260417\\
231	1931.15625\\
232	1934.979167\\
233	1930.5625\\
234	1920.234375\\
235	1929.03125\\
236	1931.802083\\
237	1926.208333\\
238	1922.796875\\
239	1920.703125\\
240	1920\\
241	1906.8\\
242	1918.010\\
243	1928.70312\\
244	1921.541667\\
245	1913.3125\\
246	1917.140625\\
247	1935.770833\\
248	1931.416667\\
249	1938.770833\\
250	1921.260417\\
251	1912.484375\\
252	1923.197917\\
253	1923.489583\\
254	1930.244792\\
255	1930.65625\\
256	1927\\
257	1919.1\\
258	1920.8281\\
259	1912.854167\\
260	1918.75\\
261	1910.5312\\
262	1912.1875\\
263	1918.458333\\
264	1922.614583\\
265	1922.197917\\
266	1926.416667\\
267	1919.916667\\
268	1927.473958\\
269	1929.072917\\
270	1922.333333\\
271	1923.286458\\
272	1923.072917\\
273	1933.869792\\
274	1929.848958\\
275	1931.671875\\
276	1934.609375\\
277	1917.078125\\
278	1907.989583\\
279	1907.078125\\
280	1905.640625\\
281	1911.135417\\
282	1905.59375\\
283	1906.526042\\
284	1915.4375\\
285	1921.994792\\
286	1924.453125\\
287	1923.291667\\
288	1933.588542\\
289	1935.473958\\
290	1927.182292\\
291	1926.770833\\
292	1926.114583\\
293	1922.609375\\
294	1923.34375\\
295	1930.536458\\
296	1933.369792\\
297	1934.729167\\
298	1925.234375\\
299	1932.307292\\
300	1926.947917\\
};
\end{axis}

\end{tikzpicture}%
}
        \caption{Average number of active
indices as a function of epochs.}
        \label{fg:nonZeroEpoc} 
    \end{minipage}
    \begin{minipage}[b]{0.45\linewidth}
 \scalebox{0.45}{\begin{tikzpicture}
\definecolor{mycolor1}{rgb}{0.63529,0.07843,0.18431}%
\definecolor{mycolor2}{rgb}{0.00000,0.44706,0.74118}%
\definecolor{mycolor3}{rgb}{0.00000,0.49804,0.00000}%
\definecolor{mycolor4}{rgb}{0.87059,0.49020,0.00000}%
\definecolor{mycolor5}{rgb}{0.00000,0.44700,0.74100}%
\definecolor{mycolor6}{rgb}{0.74902,0.00000,0.74902}%
\definecolor{mycolor7}{rgb}{0.502,0.2000,0.5902}

\begin{axis}[
font=\footnotesize,
width=7cm,
height=5.5cm,
scale only axis,
xmin=1,
xmax=300,
xtick = {0,50,...,300},
xlabel={Epochs},
xmajorgrids,
ymin=10,
ymax=70,
ytick = {10,20,...,70},
ylabel={Average Number of Recovered Indices from $\mathcal{B}$},
ylabel near ticks,
ymajorgrids,
legend style={font=\scriptsize, at={(1,1.4)},anchor=north east, draw=black,fill=white,legend cell align=left}
]

\addplot [color=mycolor1,solid,line width=1.5pt]
  table[row sep=crcr]{
1	25.94791667	\\	
2	28.05729167	\\	
3	19.82291667	\\	
4	17.22395833	\\	
5	17.24479167	\\	
6	16.671875	\\	
7	16.74479167	\\	
8	16.625	\\	
9	16	\\	
10	16.16666667	\\	
11	16.72916667	\\	
12	16.09375	\\	
13	16.85416667	\\	
14	16.6875	\\	
15	17.578125	\\	
16	17.71354167	\\	
17	16.63020833	\\	
18	17.1875	\\	
19	17.65104167	\\	
20	18.80729167	\\	
21	17.72916667	\\	
22	17.8125	\\	
23	17.55208333	\\	
24	18.390625	\\	
25	18.25	\\	
26	17.79166667	\\	
27	18.36458333	\\	
28	17.75	\\	
29	18.30208333	\\	
30	19.05729167	\\	
31	19.10416667	\\	
32	18.97916667	\\	
33	19.234375	\\	
34	19.11458333	\\	
35	19.94791667	\\	
36	19.765625	\\	
37	20.11979167	\\	
38	24.92708333	\\	
39	23.640625	\\	
40	21.375	\\	
41	21.74479167	\\	
42	22.359375	\\	
43	23.359375	\\	
44	23.77083333	\\	
45	23.90104167	\\	
46	24.14583333	\\	
47	29.02083333	\\	
48	25.93229167	\\	
49	26.39583333	\\	
50	25.796875	\\	
51	25.40104167	\\	
52	25.11979167	\\	
53	26.55729167	\\	
54	27.36979167	\\	
55	28.27083333	\\	
56	27.64583333	\\	
57	28.453125	\\	
58	29.359375	\\	
59	30.41666667	\\	
60	31.30208333	\\	
61	29.93229167	\\	
62	31	\\	
63	30.109375	\\	
64	30.59895833	\\	
65	31.14583333	\\	
66	29.91145833	\\	
67	30.50520833	\\	
68	35	\\	
69	32.75	\\	
70	32.17708333	\\	
71	31.83333333	\\	
72	31.28645833	\\	
73	37.69791667	\\	
74	35.72916667	\\	
75	33.82291667	\\	
76	34.93229167	\\	
77	34.375	\\	
78	36.421875	\\	
79	37.875	\\	
80	39.56770833	\\	
81	38.38541667	\\	
82	38.296875	\\	
83	39.81770833	\\	
84	40.04166667	\\	
85	37.63541667	\\	
86	37.625	\\	
87	35.6875	\\	
88	37.078125	\\	
89	36.59375	\\	
90	36.171875	\\	
91	38.765625	\\	
92	38.39583333	\\	
93	43.671875	\\	
94	40.39583333	\\	
95	39.81770833	\\	
96	41.19270833	\\	
97	46.5	\\	
98	44.56770833	\\	
99	43.421875	\\	
100	45.05208333	\\	
101	44.93229167	\\	
102	45.328125	\\	
103	44.86979167	\\	
104	42.84375	\\	
105	43.40104167	\\	
106	44.53645833	\\	
107	45.859375	\\	
108	45.109375	\\	
109	46.83333333	\\	
110	45.77604167	\\	
111	46.84375	\\	
112	47.08333333	\\	
113	45.46875	\\	
114	45.55208333	\\	
115	45.08333333	\\	
116	47.75520833	\\	
117	47.84895833	\\	
118	48.27083333	\\	
119	47.28125	\\	
120	46.04166667	\\	
121	45.58333333	\\	
122	46.31770833	\\	
123	48.70833333	\\	
124	47.890625	\\	
125	45.125	\\	
126	46.48958333	\\	
127	48.21354167	\\	
128	47.265625	\\	
129	46.54166667	\\	
130	45.9375	\\	
131	48.08333333	\\	
132	49.30208333	\\	
133	47.8125	\\	
134	48.63020833	\\	
135	47.50520833	\\	
136	51.11458333	\\	
137	52.22916667	\\	
138	50.703125	\\	
139	49.453125	\\	
140	49.18229167	\\	
141	49.50520833	\\	
142	50.60416667	\\	
143	51.36979167	\\	
144	51.18229167	\\	
145	49.82291667	\\	
146	52.78125	\\	
147	50.953125	\\	
148	49.203125	\\	
149	51.53645833	\\	
150	50.58854167	\\	
151	51.49479167	\\	
152	52.55729167	\\	
153	53.08333333	\\	
154	54.75520833	\\	
155	53.3125	\\	
156	54.734375	\\	
157	57.015625	\\	
158	57.49479167	\\	
159	58.31770833	\\	
160	56.44270833	\\	
161	54.42708333	\\	
162	56.5	\\	
163	53.13541667	\\	
164	52.64583333	\\	
165	52.28125	\\	
166	51.96354167	\\	
167	51.55208333	\\	
168	50.65625	\\	
169	52.15104167	\\	
170	53.65104167	\\	
171	53.0625	\\	
172	50.74479167	\\	
173	52.70833333	\\	
174	52.96354167	\\	
175	54.19791667	\\	
176	54.56770833	\\	
177	54.671875	\\	
178	55.63541667	\\	
179	55.38541667	\\	
180	55.94270833	\\	
181	55.19270833	\\	
182	55.03125	\\	
183	53.44270833	\\	
184	54.93229167	\\	
185	54.55208333	\\	
186	51.83854167	\\	
187	54.36979167	\\	
188	55.984375	\\	
189	55.484375	\\	
190	57.72916667	\\	
191	56.65625	\\	
192	55.02604167	\\	
193	53.9375	\\	
194	55.22916667	\\	
195	55.94791667	\\	
196	55.984375	\\	
197	55.5625	\\	
198	53.53125	\\	
199	51.84375	\\	
200	57.125	\\	
201	57.9375	\\	
202	58.45833333	\\	
203	57.76041667	\\	
204	55.65104167	\\	
205	57.00520833	\\	
206	56.00520833	\\	
207	55.484375	\\	
208	55.875	\\	
209	59.16145833	\\	
210	60.08854167	\\	
211	60.89583333	\\	
212	59.23958333	\\	
213	59.98958333	\\	
214	61.08333333	\\	
215	61.41666667	\\	
216	62.99479167	\\	
217	63.046875	\\	
218	62.640625	\\	
219	63.44270833	\\	
220	62.796875	\\	
221	62.00520833	\\	
222	61.41145833	\\	
223	59.921875	\\	
224	58.390625	\\	
225	59.55208333	\\	
226	58.27604167	\\	
227	58.40625	\\	
228	59.54166667	\\	
229	59.97916667	\\	
230	60.77604167	\\	
231	61.32291667	\\	
232	62.20833333	\\	
233	62.28125	\\	
234	60.453125	\\	
235	61.578125	\\	
236	61.73958333	\\	
237	62.75	\\	
238	61.359375	\\	
239	61.40625	\\	
240	60.515625	\\	
241	59.17708333	\\	
242	60.72916667	\\	
243	62.75520833	\\	
244	61.59895833	\\	
245	61.14583333	\\	
246	61.04166667	\\	
247	64.49479167	\\	
248	62.609375	\\	
249	63.20833333	\\	
250	62.80729167	\\	
251	59.80208333	\\	
252	62.921875	\\	
253	61.68229167	\\	
254	63	\\	
255	63.203125	\\	
256	61.86458333	\\	
257	60.515625	\\	
258	61.796875	\\	
259	61.06770833	\\	
260	59.97395833	\\	
261	59.421875	\\	
262	61.02604167	\\	
263	61.75	\\	
264	62.48958333	\\	
265	62.33854167	\\	
266	63.30729167	\\	
267	61.17708333	\\	
268	62.19270833	\\	
269	63.65104167	\\	
270	61.63020833	\\	
271	62.30729167	\\	
272	61.17708333	\\	
273	63.34895833	\\	
274	63.23958333	\\	
275	63.84895833	\\	
276	64.5	\\	
277	61.734375	\\	
278	60.671875	\\	
279	60.30208333	\\	
280	59.50520833	\\	
281	60.83854167	\\	
282	60.46875	\\	
283	60.59895833	\\	
284	61.69791667	\\	
285	62.28645833	\\	
286	63.38541667	\\	
287	62.80208333	\\	
288	64.07291667	\\	
289	65.36979167	\\	
290	64.25520833	\\	
291	64.15625	\\	
292	64.71875	\\	
293	62.58333333	\\	
294	62.86979167	\\	
295	63.25	\\	
296	64.5	\\	
297	65.67708333	\\	
298	63.92708333	\\	
299	65.52083333	\\	
300	63.890625	\\	
};
\end{axis}

\end{tikzpicture}%
}
        \caption{Average Number of Recovered Indices from $\mathcal{B}$}
        \label{fg:actiFA} 
    \end{minipage}
\end{figure}

\begin{figure}
    \centering
    \begin{minipage}[b]{0.45\linewidth}
 \scalebox{0.45}{\begin{tikzpicture}
\definecolor{mycolor1}{rgb}{0.63529,0.07843,0.18431}%
\definecolor{mycolor2}{rgb}{0.00000,0.44706,0.74118}%
\definecolor{mycolor3}{rgb}{0.00000,0.49804,0.00000}%
\definecolor{mycolor4}{rgb}{0.87059,0.49020,0.00000}%
\definecolor{mycolor5}{rgb}{0.00000,0.44700,0.74100}%
\definecolor{mycolor6}{rgb}{0.74902,0.00000,0.74902}%
\definecolor{mycolor7}{rgb}{0.502,0.2000,0.5902}

\begin{axis}[
font=\footnotesize,
width=7cm,
height=5.5cm,
scale only axis,
xmin=1,
xmax=300,
xtick = {0,50,...,300},
xlabel={Epochs},
xmajorgrids,
ymin=0,
ymax=1,
ytick = {0,0.2,...,1},
ylabel={Average Probability of Detection},
ylabel near ticks,
ymajorgrids,
legend style={font=\scriptsize, at={(1,1.4)},anchor=north east, draw=black,fill=white,legend cell align=left}
]

\addplot [color=mycolor1,solid,line width=1.5pt]
  table[row sep=crcr]{
1	0.738946759	\\
2	0.569772377	\\
3	0.519675926	\\
4	0.516917438	\\
5	0.50929784	\\
6	0.505979938	\\
7	0.516010802	\\
8	0.514293981	\\
9	0.519579475	\\
10	0.523167438	\\
11	0.527121914	\\
12	0.4890625	\\
13	0.473804012	\\
14	0.448958333	\\
15	0.433526235	\\
16	0.430073302	\\
17	0.423996914	\\
18	0.422974537	\\
19	0.422048611	\\
20	0.391512346	\\
21	0.418904321	\\
22	0.402256944	\\
23	0.401774691	\\
24	0.390219907	\\
25	0.390104167	\\
26	0.402083333	\\
27	0.40003858	\\
28	0.399807099	\\
29	0.37507716	\\
30	0.348302469	\\
31	0.347299383	\\
32	0.341300154	\\
33	0.341473765	\\
34	0.329822531	\\
35	0.323881173	\\
36	0.331674383	\\
37	0.306655093	\\
38	0.357523148	\\
39	0.379378858	\\
40	0.388387346	\\
41	0.376292438	\\
42	0.368402778	\\
43	0.364544753	\\
44	0.363676698	\\
45	0.34681713	\\
46	0.357966821	\\
47	0.399074074	\\
48	0.425462963	\\
49	0.41244213	\\
50	0.426851852	\\
51	0.428722994	\\
52	0.434008488	\\
53	0.417283951	\\
54	0.397106481	\\
55	0.388406636	\\
56	0.404166667	\\
57	0.394367284	\\
58	0.373707562	\\
59	0.353973765	\\
60	0.343094136	\\
61	0.344560185	\\
62	0.342341821	\\
63	0.363155864	\\
64	0.361496914	\\
65	0.346161265	\\
66	0.345177469	\\
67	0.339834105	\\
68	0.378684414	\\
69	0.414872685	\\
70	0.418904321	\\
71	0.417920525	\\
72	0.408757716	\\
73	0.419174383	\\
74	0.458564815	\\
75	0.473977623	\\
76	0.459780093	\\
77	0.447183642	\\
78	0.427411265	\\
79	0.405266204	\\
80	0.40304784	\\
81	0.42341821	\\
82	0.411168981	\\
83	0.401716821	\\
84	0.406500772	\\
85	0.434876543	\\
86	0.44222608	\\
87	0.4609375	\\
88	0.455767747	\\
89	0.455015432	\\
90	0.44068287	\\
91	0.411516204	\\
92	0.405941358	\\
93	0.436959877	\\
94	0.455536265	\\
95	0.460628858	\\
96	0.44068287	\\
97	0.452334105	\\
98	0.468537809	\\
99	0.493016975	\\
100	0.486516204	\\
101	0.492727623	\\
102	0.478742284	\\
103	0.477854938	\\
104	0.511593364	\\
105	0.496315586	\\
106	0.490740741	\\
107	0.486168981	\\
108	0.477797068	\\
109	0.465798611	\\
110	0.478742284	\\
111	0.466396605	\\
112	0.480150463	\\
113	0.500945216	\\
114	0.499652778	\\
115	0.501929012	\\
116	0.486631944	\\
117	0.489776235	\\
118	0.477256944	\\
119	0.482253086	\\
120	0.493576389	\\
121	0.493209877	\\
122	0.495505401	\\
123	0.461844136	\\
124	0.461304012	\\
125	0.492708333	\\
126	0.482445988	\\
127	0.457118056	\\
128	0.485513117	\\
129	0.483082562	\\
130	0.473861883	\\
131	0.460821759	\\
132	0.442689043	\\
133	0.450212191	\\
134	0.451311728	\\
135	0.452719907	\\
136	0.498321759	\\
137	0.496045525	\\
138	0.506886574	\\
139	0.520216049	\\
140	0.524479167	\\
141	0.519367284	\\
142	0.499421296	\\
143	0.50150463	\\
144	0.515625	\\
145	0.522434414	\\
146	0.490277778	\\
147	0.504089506	\\
148	0.522569444	\\
149	0.502314815	\\
150	0.495852623	\\
151	0.498148148	\\
152	0.486111111	\\
153	0.485648148	\\
154	0.464467593	\\
155	0.477237654	\\
156	0.472704475	\\
157	0.444193673	\\
158	0.442824074	\\
159	0.430227623	\\
160	0.447106481	\\
161	0.460088735	\\
162	0.455478395	\\
163	0.488946759	\\
164	0.49992284	\\
165	0.497241512	\\
166	0.500868056	\\
167	0.507118056	\\
168	0.515316358	\\
169	0.508140432	\\
170	0.489699074	\\
171	0.50152392	\\
172	0.503722994	\\
173	0.494058642	\\
174	0.48900463	\\
175	0.482156636	\\
176	0.490740741	\\
177	0.488097994	\\
178	0.467033179	\\
179	0.478530093	\\
180	0.475135031	\\
181	0.471778549	\\
182	0.473861883	\\
183	0.490933642	\\
184	0.466975309	\\
185	0.480748457	\\
186	0.500752315	\\
187	0.478375772	\\
188	0.468595679	\\
189	0.47023534	\\
190	0.448939043	\\
191	0.461998457	\\
192	0.461477623	\\
193	0.47496142	\\
194	0.469714506	\\
195	0.463464506	\\
196	0.461766975	\\
197	0.46716821	\\
198	0.478433642	\\
199	0.496759259	\\
200	0.519656636	\\
201	0.518557099	\\
202	0.505459105	\\
203	0.520196759	\\
204	0.541589506	\\
205	0.531751543	\\
206	0.525810185	\\
207	0.53599537	\\
208	0.526851852	\\
209	0.510281636	\\
210	0.49211034	\\
211	0.483931327	\\
212	0.494174383	\\
213	0.49685571	\\
214	0.478530093	\\
215	0.480787037	\\
216	0.465760031	\\
217	0.470023148	\\
218	0.464351852	\\
219	0.457716049	\\
220	0.47025463	\\
221	0.471682099	\\
222	0.478356481	\\
223	0.487326389	\\
224	0.50318287	\\
225	0.502893519	\\
226	0.508854167	\\
227	0.511940586	\\
228	0.50474537	\\
229	0.497820216	\\
230	0.489949846	\\
231	0.47650463	\\
232	0.470216049	\\
233	0.478317901	\\
234	0.491608796	\\
235	0.481790123	\\
236	0.469290123	\\
237	0.475713735	\\
238	0.487210648	\\
239	0.483641975	\\
240	0.487827932	\\
241	0.510146605	\\
242	0.491975309	\\
243	0.474672068	\\
244	0.481462191	\\
245	0.494116512	\\
246	0.489409722	\\
247	0.464756944	\\
248	0.470852623	\\
249	0.453877315	\\
250	0.476446759	\\
251	0.495987654	\\
252	0.483217593	\\
253	0.483294753	\\
254	0.472222222	\\
255	0.471180556	\\
256	0.477835648	\\
257	0.486766975	\\
258	0.475617284	\\
259	0.493364198	\\
260	0.489467593	\\
261	0.502334105	\\
262	0.492959105	\\
263	0.485050154	\\
264	0.477912809	\\
265	0.477854938	\\
266	0.470833333	\\
267	0.483622685	\\
268	0.466801698	\\
269	0.469810957	\\
270	0.485551698	\\
271	0.481925154	\\
272	0.481462191	\\
273	0.461902006	\\
274	0.465412809	\\
275	0.458641975	\\
276	0.457291667	\\
277	0.487866512	\\
278	0.503375772	\\
279	0.500906636	\\
280	0.514178241	\\
281	0.500443673	\\
282	0.505150463	\\
283	0.500424383	\\
284	0.485802469	\\
285	0.474228395	\\
286	0.476929012	\\
287	0.478260031	\\
288	0.456867284	\\
289	0.45632716	\\
290	0.46720679	\\
291	0.462712191	\\
292	0.464718364	\\
293	0.471778549	\\
294	0.474209105	\\
295	0.46251929	\\
296	0.457484568	\\
297	0.452314815	\\
298	0.467418981	\\
299	0.455806327	\\
300	0.462827932	\\
};
\end{axis}

\end{tikzpicture}%
}
        \caption{Average probability of Detection.}
        \label{fg:DE} 
    \end{minipage}
    \begin{minipage}[b]{0.45\linewidth}
 \scalebox{0.45}{\begin{tikzpicture}
\definecolor{mycolor1}{rgb}{0.63529,0.07843,0.18431}%
\definecolor{mycolor2}{rgb}{0.00000,0.44706,0.74118}%
\definecolor{mycolor3}{rgb}{0.00000,0.49804,0.00000}%
\definecolor{mycolor4}{rgb}{0.87059,0.49020,0.00000}%
\definecolor{mycolor5}{rgb}{0.00000,0.44700,0.74100}%
\definecolor{mycolor6}{rgb}{0.74902,0.00000,0.74902}%
\definecolor{mycolor7}{rgb}{0.502,0.2000,0.5902}

\begin{axis}[
font=\footnotesize,
width=7cm,
height=5.5cm,
scale only axis,
xmin=1,
xmax=300,
xtick = {0,50,...,300},
xlabel={Epochs},
xmajorgrids,
ymin=0,
ymax=1,
ytick = {0,0.2,...,1},
ylabel={Average Probability of False Alarm},
ylabel near ticks,
ymajorgrids,
legend style={font=\scriptsize, at={(1,1.4)},anchor=north east, draw=black,fill=white,legend cell align=left}
]

\addplot [color=mycolor1,solid,line width=1.5pt]
  table[row sep=crcr]{
1	0.210293653	\\
2	0.299970482	\\
3	0.312874389	\\
4	0.309202817	\\
5	0.315543114	\\
6	0.316786644	\\
7	0.310824681	\\
8	0.308447302	\\
9	0.305649036	\\
10	0.305361017	\\
11	0.303891922	\\
12	0.328230138	\\
13	0.337644192	\\
14	0.350248591	\\
15	0.367275678	\\
16	0.366177079	\\
17	0.36663763	\\
18	0.374011397	\\
19	0.376397949	\\
20	0.402773019	\\
21	0.375385169	\\
22	0.392581994	\\
23	0.396884479	\\
24	0.40173544	\\
25	0.397197854	\\
26	0.389641305	\\
27	0.393743477	\\
28	0.391780723	\\
29	0.413059737	\\
30	0.437195758	\\
31	0.437746989	\\
32	0.444214026	\\
33	0.445383006	\\
34	0.454077307	\\
35	0.454745232	\\
36	0.449422023	\\
37	0.477720852	\\
38	0.435333922	\\
39	0.423776502	\\
40	0.412469116	\\
41	0.428435688	\\
42	0.428368015	\\
43	0.427924721	\\
44	0.429327626	\\
45	0.449344929	\\
46	0.44148655	\\
47	0.409768561	\\
48	0.38068784	\\
49	0.39537402	\\
50	0.39154222	\\
51	0.385076762	\\
52	0.381429273	\\
53	0.396585466	\\
54	0.411035918	\\
55	0.418036053	\\
56	0.403536307	\\
57	0.410579374	\\
58	0.433118356	\\
59	0.447065805	\\
60	0.457459071	\\
61	0.452357298	\\
62	0.455908401	\\
63	0.439847479	\\
64	0.437500297	\\
65	0.45458754	\\
66	0.457746232	\\
67	0.465779151	\\
68	0.438152608	\\
69	0.401400661	\\
70	0.402146994	\\
71	0.408436956	\\
72	0.420321718	\\
73	0.408670535	\\
74	0.379282257	\\
75	0.366818822	\\
76	0.380126949	\\
77	0.394660421	\\
78	0.408725678	\\
79	0.421450572	\\
80	0.425161997	\\
81	0.402902073	\\
82	0.416169039	\\
83	0.423115906	\\
84	0.424548924	\\
85	0.391972326	\\
86	0.390463729	\\
87	0.376874623	\\
88	0.376700081	\\
89	0.376715606	\\
90	0.389173454	\\
91	0.416096356	\\
92	0.424011224	\\
93	0.408829511	\\
94	0.393466853	\\
95	0.394558776	\\
96	0.407773674	\\
97	0.400285542	\\
98	0.386754385	\\
99	0.367878796	\\
100	0.371859503	\\
101	0.367764732	\\
102	0.380303967	\\
103	0.377728536	\\
104	0.351877054	\\
105	0.358827678	\\
106	0.3641691	\\
107	0.37166392	\\
108	0.37849137	\\
109	0.39007702	\\
110	0.37732735	\\
111	0.387292515	\\
112	0.374301331	\\
113	0.361554174	\\
114	0.360993089	\\
115	0.359884997	\\
116	0.37348125	\\
117	0.370072471	\\
118	0.377473602	\\
119	0.370869663	\\
120	0.364963819	\\
121	0.362416057	\\
122	0.365569093	\\
123	0.393958524	\\
124	0.390841793	\\
125	0.365232435	\\
126	0.377088024	\\
127	0.394716313	\\
128	0.374163052	\\
129	0.375571525	\\
130	0.385019808	\\
131	0.397516404	\\
132	0.407887797	\\
133	0.398228172	\\
134	0.400741224	\\
135	0.401879864	\\
136	0.368818298	\\
137	0.373054281	\\
138	0.366228462	\\
139	0.354545922	\\
140	0.351373124	\\
141	0.354532159	\\
142	0.368864293	\\
143	0.369326645	\\
144	0.359140406	\\
145	0.351679623	\\
146	0.378150426	\\
147	0.364552408	\\
148	0.352063398	\\
149	0.370879527	\\
150	0.374580945	\\
151	0.374077651	\\
152	0.385472839	\\
153	0.382946268	\\
154	0.395440767	\\
155	0.38835226	\\
156	0.390023018	\\
157	0.414860631	\\
158	0.419407342	\\
159	0.427637095	\\
160	0.415541616	\\
161	0.40317447	\\
162	0.407059543	\\
163	0.380883826	\\
164	0.370939879	\\
165	0.373417903	\\
166	0.368487768	\\
167	0.364386124	\\
168	0.358541261	\\
169	0.364407206	\\
170	0.379338622	\\
171	0.371059161	\\
172	0.363500203	\\
173	0.372403225	\\
174	0.380013178	\\
175	0.383567711	\\
176	0.382660668	\\
177	0.385935863	\\
178	0.396878273	\\
179	0.392399297	\\
180	0.3932847	\\
181	0.39324163	\\
182	0.390631918	\\
183	0.382953206	\\
184	0.397505017	\\
185	0.388011024	\\
186	0.368580883	\\
187	0.386465539	\\
188	0.397445352	\\
189	0.394983204	\\
190	0.412957846	\\
191	0.401633826	\\
192	0.397386285	\\
193	0.389912914	\\
194	0.394513959	\\
195	0.404070711	\\
196	0.408362595	\\
197	0.402934095	\\
198	0.393054524	\\
199	0.378175352	\\
200	0.367528323	\\
201	0.365358175	\\
202	0.375618756	\\
203	0.364515719	\\
204	0.347097302	\\
205	0.354322581	\\
206	0.358210916	\\
207	0.352082357	\\
208	0.359498446	\\
209	0.376492266	\\
210	0.388184428	\\
211	0.396048913	\\
212	0.385823571	\\
213	0.386141496	\\
214	0.396886651	\\
215	0.397240002	\\
216	0.406241819	\\
217	0.406523199	\\
218	0.409169875	\\
219	0.413462449	\\
220	0.405905024	\\
221	0.404866083	\\
222	0.399070122	\\
223	0.388913814	\\
224	0.376926415	\\
225	0.378168245	\\
226	0.373384779	\\
227	0.371154221	\\
228	0.378366508	\\
229	0.384024838	\\
230	0.389514462	\\
231	0.399057098	\\
232	0.404556413	\\
233	0.398267617	\\
234	0.38657235	\\
235	0.398774294	\\
236	0.403776522	\\
237	0.402258842	\\
238	0.393030842	\\
239	0.396822963	\\
240	0.393379786	\\
241	0.37302345	\\
242	0.388004269	\\
243	0.404591512	\\
244	0.398700717	\\
245	0.391565974	\\
246	0.392834425	\\
247	0.41167704	\\
248	0.405248987	\\
249	0.417503814	\\
250	0.403658526	\\
251	0.384457301	\\
252	0.398586149	\\
253	0.396642632	\\
254	0.405600759	\\
255	0.404529476	\\
256	0.397601388	\\
257	0.391958796	\\
258	0.4048279	\\
259	0.387715703	\\
260	0.387990674	\\
261	0.377447422	\\
262	0.390119597	\\
263	0.39731172	\\
264	0.400534888	\\
265	0.400040836	\\
266	0.407825414	\\
267	0.393830714	\\
268	0.407488068	\\
269	0.405868148	\\
270	0.391729352	\\
271	0.39610909	\\
272	0.392968921	\\
273	0.411448408	\\
274	0.411212947	\\
275	0.415792812	\\
276	0.421249892	\\
277	0.395270439	\\
278	0.379640281	\\
279	0.3796905	\\
280	0.370405477	\\
281	0.383356147	\\
282	0.38031572	\\
283	0.382124268	\\
284	0.395241822	\\
285	0.40467921	\\
286	0.403060247	\\
287	0.400948999	\\
288	0.416741102	\\
289	0.417762744	\\
290	0.412232543	\\
291	0.414833105	\\
292	0.415388156	\\
293	0.405683637	\\
294	0.404050114	\\
295	0.409683747	\\
296	0.413695946	\\
297	0.425670925	\\
298	0.412358222	\\
299	0.421222097	\\
300	0.414061473	\\
};
\end{axis}

\end{tikzpicture}%
}
        \caption{Average probability of False Alarm.}
        \label{fg:FA} 
    \end{minipage}
\end{figure}
\vspace{-5mm}

\section{Summary}
This article demonstrates that multiple access techniques can be leveraged to construct CS algorithms that can compress gradient updates for models that have thousands of parameters.
This approach enables over-the-air FL at scale.
Secondly, we offer a methodology based on how the gradient vectors change over epochs to give insight into parameter selection for CS-based, over-the-air FL problems.

\label{Bibliography}
\bibliographystyle{IEEEbib}
\bibliography{referencesFL}
\end{document}